\PassOptionsToPackage{dvipsnames}{xcolor}
\documentclass[usenatbib]{mnras}
\pdfoutput=1

\usepackage{graphicx}
\usepackage{amsmath,amssymb,amstext}
\usepackage[T1]{fontenc}
\usepackage{txfonts}
\usepackage{booktabs}
\usepackage{hyperref}%
\usepackage{hypernat}%
\usepackage[all]{hypcap}
\usepackage{url}
\usepackage{xcolor}
\usepackage[capitalise]{cleveref}
\crefname{equation}{Eq.}{Eqs.}
\Crefname{equation}{Eq.}{Eqs.}
\crefformat{equation}{#2Eq.~#1#3}
\Crefformat{equation}{#2Eq.~#1#3}
\crefmultiformat{equation}{Eqs.~#2#1#3}{ and~#2#1#3}{, #2#1#3}{ and~#2#1#3}
\Crefmultiformat{equation}{Eqs.~#2#1#3}{ and~#2#1#3}{, #2#1#3}{ and~#2#1#3}
\usepackage{orcidlink}

\newcommand{\alphaSED}{\alpha_{\rm SED}}
\newcommand{\alphaTol}{\alpha_{\rm Tol}}
\newcommand{\alphaDDR}{\alpha_{\rm DDR}}
\newcommand{\alphaTolMean}{\alpha_{{\rm Tol},\rm mean}}
\newcommand{\alphaTolMax}{\alpha_{{\rm Tol},\rm max}}
\newcommand{\alphaTolModalMean}{\alpha_{{\rm Tol},\rm modal\,mean}}
\newcommand{\alphaTolModalMax}{\alpha_{{\rm Tol},\rm modal\,max}}
\newcommand{\Isb}{I_{\rm SB}}
\newcommand{\Isbobs}{I_{\rm SB}^{\rm obs}}

\newcommand{\DLum}{D_{L}}
\newcommand{\DAng}{D_{A}}
\newcommand{\rhoL}{\rho_{L}}
\newcommand{\zspec}{z_\text{spec}}

\title[TNG meets the Tolman test]
{Forward-modelling the Tolman and distance-duality tests with \texttt{IllustrisTNG}}

\author[Desmond et al.]{
Harry Desmond\orcidlink{0000-0003-0685-9791}$^{1}$\thanks{E-mail: \href{mailto:harry.desmond@port.ac.uk}{harry.desmond@port.ac.uk}},
Tariq Yasin$^{2}$, Richard Stiskalek\orcidlink{0000-0002-0986-314X}$^{2}$ and Sebastian von Hausegger\orcidlink{0000-0002-6274-1424}$^{3}$\\
$^{1}$Institute of Cosmology \& Gravitation, University of Portsmouth, Portsmouth PO1 3FX, UK\\
$^{2}$Astrophysics, University of Oxford, Denys Wilkinson Building, Keble Road, Oxford, OX1 3RH, UK\\
$^{3}$Rudolf Peierls Centre for Theoretical Physics, University of Oxford, Parks Road, Oxford OX1 3PU, UK
}

\pubyear{2026}

\begin{document}
\label{firstpage}
\pagerange{\pageref{firstpage}--\pageref{lastpage}}
\maketitle

\begin{abstract}
The Tolman surface-brightness test and the angular-size distance-duality test are two complementary probes of the same underlying relation between luminosity and angular-diameter distance, $\DLum=(1+z)^2\DAng$, as holds in any metric theory of gravity where photon number is conserved. Both tests have recently delivered \textit{a priori} surprising signals: JWST/ASTRODEEP measurements yield a surface brightness scaling with redshift much flatter than the expected value,
and ultracompact radio sources also appear to follow a flatter $D_L/D_A$ scaling with redshift. These results have been suggested to support non-expanding cosmologies; however, they are also sensitive to astrophysical and instrumental effects.
We test whether these results indicate genuine departures from standard cosmology by forward-modelling observed surface-brightness evolution in the \texttt{IllustrisTNG} cosmological hydrodynamical simulation, with an empirical mock-spectroscopic selection trained on ASTRODEEP.
We show that the astrophysical evolution relevant for both tests may be effectively parametrised as a single power-law exponent for the luminosity density as a function of redshift, for which the simulation gives $\gamma=2.23\pm0.20$ across realistic aperture conventions.
This value
is approximately sufficient
to explain both the Tolman and distance-duality signals within standard cosmology and galaxy formation physics, with a small discrepancy for the latter suggesting that radio AGN evolve slightly more strongly than bright galaxies.
\end{abstract}

\begin{keywords}
cosmology: observations -- galaxies: high-redshift -- galaxies: photometry -- methods: numerical
\end{keywords}

\section{Introduction}\label{sec:intro}

The redshift--distance relation is a fundamental relation in cosmology. In any metric theory of gravity that conserves photon number along null geodesics, luminosity and angular-diameter distances are linked by the Etherington relation $\DLum=(1+z)^2\DAng$~\citep{Etherington1933}. This relation is geometric in origin and so is independent of the gravitational field equations and hence the dynamics of the expansion; it can therefore be used as an unambiguous test that the Universe permits a metric description~\citep{Ellis2007}.

Two operational tests of this prediction have a long history. The first is the Tolman surface-brightness (SB) test~\citep{Tolman1930,Tolman1934}: for a non-evolving extended source the bolometric SB should fall as $\Isb\propto(1+z)^{-4}$, while a measurement at fixed observer-frame frequency in a single broad band gives $\Isbobs\propto(1+z)^{-(3+\alphaSED)}$ with $\alphaSED$ the rest-frame spectral energy distribution (SED) slope of the source. The second is the angular-size--redshift relation for non-evolving compact sources \citep{Kapahi1987,Gurvits1994,GurvitsKellermannFrey1999,JacksonJannetta2006,CaoBiesiada2017}: an object of fixed rest-frame physical size has a redshift-dependent angular size that, combined with a flux measurement, can be reduced to the duality ratio $\DLum/\DAng$ (hereafter the distance-duality relation, DDR).
Cluster Sunyaev--Zel'dovich measurements \citep[e.g.][]{Holanda2010} and %
strong-lensing Einstein-radius and velocity-dispersion measurements \citep[e.g.][]{Liao2016} provide further independent handles (see also \citealt{BassettKunz2004} and \citealt{Uzan2004} for early introductions to these tests).

Both tests have very recently produced results seemingly threatening to standard cosmology. \citet{TolmanJWST} report observed-frame Tolman redshift scaling of $\Isb\propto(1+z)^{-1.3}$
from the ASTRODEEP-JWST catalogue \citep{Merlin2024}, which they compare primarily with the bolometric Tolman expectation $\Isb\propto(1+z)^{-4}$. Because our forward model works with observed broad-band flux densities, we use below the corresponding specific-flux prediction $\Isbobs\propto(1+z)^{-(3+\alphaSED)}$; for our fiducial $\alphaSED=0.3$ this is still much steeper than the measured slope.\footnote{In fact,~\citet{TolmanJWST} also work with a broad-band flux, making the relevant reference slope $-3-\alpha_\text{SED}$ rather than $-4$. We return to this issue in Sec.~\ref{sec:discussion}.}
\citet{Li2023} combines the JJ2.29 \citep{JacksonJannetta2006} and GKF5.0 \citep{GurvitsKellermannFrey1999} ultracompact-radio samples to construct an empirical $\DLum/\DAng(z)$, and finds that this closely matches
$\DLum=(1+z)\DAng$.
Such findings could be framed as evidence for a non-expanding cosmology in which tired light provides the mechanism to generate redshifts~\citep[e.g.][]{Lerner2018,LopezCorredoira2015}; for simplicity we refer to these simply as ``non-expanding'' or ``non-exp.'' models hereafter. Alternatively, they may also be indicative of strong evolution of the source populations within standard Friedmann--Robertson--Walker (FRW) cosmology, or simply as artefacts of photometric systematics. Source evolution has been considered in the classical Tolman setting, where the measured passive luminosity evolution of, e.g., cluster ellipticals accounts for the shortfall of the recovered exponent from its bolometric value \citep{Pahre1996,LubinSandage2001d}, and has been revisited recently in a JWST-era volume Tolman test \citep{Conselice2026}. What has not been done is a forward-model calibration of the intrinsic-evolution exponent itself in a Lambda cold dark matter ($\Lambda$CDM) galaxy-formation simulation, propagated through both the surface-brightness and the distance-duality inferences jointly. That is the goal of this paper.

It is important to realise that while the fundamental redshift scaling $\DLum=(1+z)^{2}\DAng$ arises purely from metric geometry, \emph{measuring} either distance through a population of sources and hence reconstructing this relation empirically requires a model for the intrinsic luminosity (in the case of $\DLum$) and intrinsic physical size (in the case of $\DAng$) of those sources at each redshift. Unless dealing with a standard candle or ruler this has to be done by modelling the redshift evolution of the population, and
any unmodelled redshift evolution in the population's intrinsic luminosity or size can bias the inferred $\DLum$ or $\DAng$.
This is the mechanism by which a population of sources can \emph{appear} to fail the distance-duality test while in fact obeying the conventional relation, as has been appreciated in the literature (e.g.~\citealt{LopezCorredoira2015,Lerner2018}). It is in fact the mechanism we find to be at play. %

Although the Tolman surface brightness test and the angular-size distance duality test both depend on the sources' $\DLum$ and $\DAng$, they are conventionally treated as independent.  Indeed they do respond to different intrinsic quantities: the Tolman observable measures projected luminosity density (with $A_\text{phys}$ the rest-frame source area on the sky), $\Sigma_L\equiv L/A_{\rm phys}$, whereas the \citet{Li2023} distance-duality reduction uses volumetric luminosity density, $\rhoL\equiv L/V$ (with $V$ the rest-frame source volume). Their redshift-evolution exponents therefore need not coincide \textit{a priori}. A central question for the forward model is therefore whether a realistic $\Lambda$CDM galaxy population nevertheless links these two evolutions strongly enough to place the tests on a common physical footing, which we answer here in the affirmative.

\begin{figure}
\centering
\includegraphics[width=\columnwidth]{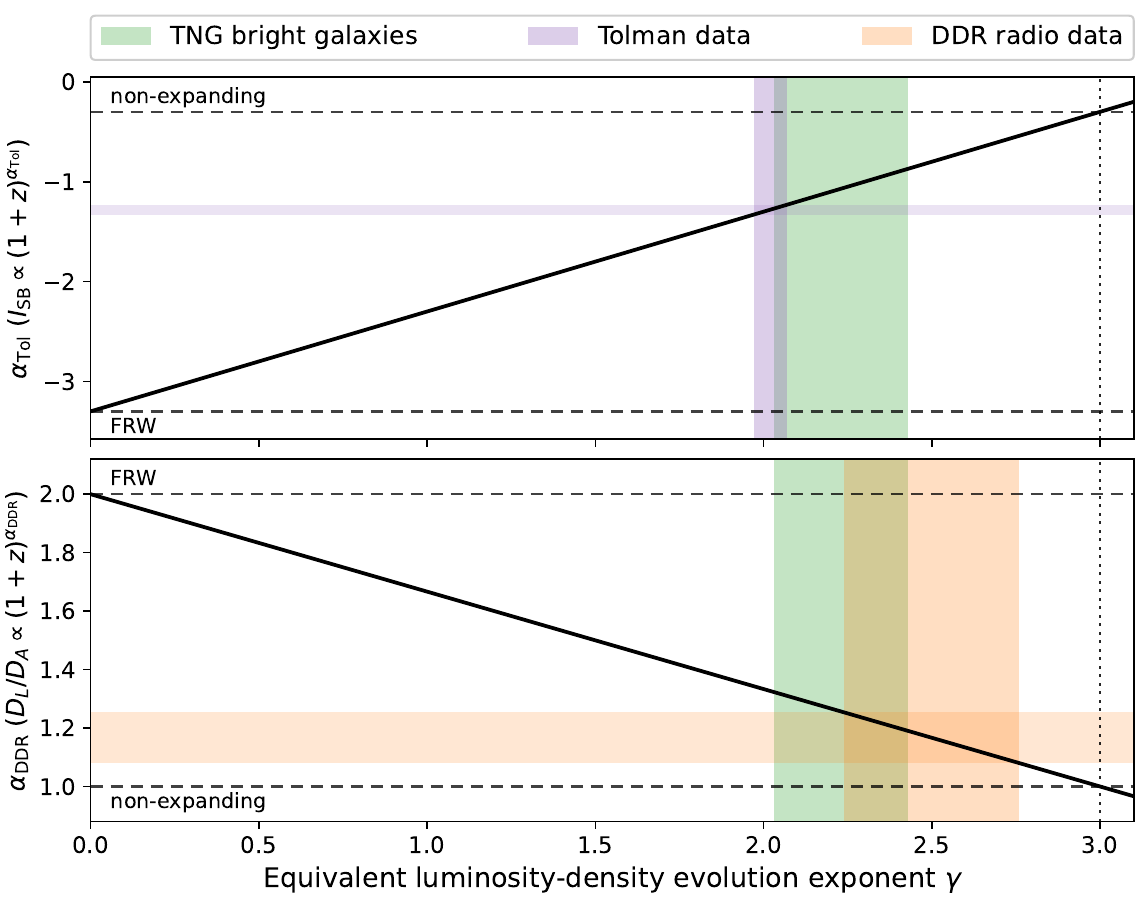}
\caption{Overview of the paper's main result. The black lines show how the Tolman exponent $\alphaTol$ and the DDR exponent $\alphaDDR$ map onto the equivalent luminosity-density evolution exponent $\gamma$ needed to generate the redshift trends within FRW cosmology. The purple and orange horizontal bands show the Tolman and DDR redshift evolution measured in the data, while the vertical bands show the corresponding luminosity-density evolution that would be required to generate these observations in an FRW universe. The green band shows the luminosity-density evolution measured in the TNG cosmological simulation. This overlaps with both mapped data bands, suggesting that source-evolution effects within $\Lambda$CDM are sufficient to explain the relevant observations without invoking alternative cosmologies.}
\label{fig:gamma_schematic}
\end{figure}

To test whether the intrinsic evolution necessary to reproduce the Tolman and distance-duality data is an expectation of $\Lambda$CDM we interrogate the state-of-the-art \texttt{IllustrisTNG-100} simulation (hereafter TNG) \citep{Springel2018,Pillepich2018,Nelson2018,Naiman2018,Marinacci2018}, taking mock HST and JWST imaging from the MAST archive \citep{NelsonHLSP,Vogelsberger2020,Snyder2022} and applying a single empirical spectroscopic-selection model trained on ASTRODEEP. We measure the Tolman and DDR-relevant evolution terms separately in the simulated bright-galaxy sample, and then ask whether their relationship explains the observational results.
This tests whether source evolution, fully within the context of standard $\Lambda$CDM galaxy formation, can account for the broad behaviour of both measurements; any residual mismatch with the compact-radio DDR result may then indicate a difference between the modelled bright galaxy population and unmodelled radio AGN population. Our principal result, summarised in Fig.~\ref{fig:gamma_schematic}, is that the root cause of both Tolman and DDR measurements is a roughly power-law evolution of source luminosity density with redshift, with a value for the exponent agreeing between the datasets and with the TNG expectation.

The structure of the paper is as follows. Sec.~\ref{sec:data} describes the observed and simulated data used. Sec.~\ref{sec:method} sets out the methodology, including the projected and volumetric intrinsic-evolution terms, the slope decomposition and the empirical mock-spectroscopic selection. Sec.~\ref{sec:results} documents our results: reproduction of the observational measurements, the TNG unification result and forward-model variants and the photometric systematics budget. Sec.~\ref{sec:discussion} interprets the results and Sec.~\ref{sec:conclusion} concludes. Throughout the paper, log has base 10.
The important quantities for our analysis, that recur across sections, are collected in Table~\ref{tab:notation}.

\section{Observed and simulated data}\label{sec:data}

\subsection{JWST-ASTRODEEP and ultra-compact radio sources}\label{ssec:obs_data}

\begin{table}
	\centering
	\caption{Notation used throughout the paper.}
	\label{tab:notation}
	\begin{tabular}{lp{0.70\columnwidth}}
		\toprule
		Symbol & Meaning \\
		\midrule
		$\gamma_\Sigma$ & Projected evolution exponent: $\Sigma_L\propto(1+z)^{\gamma_\Sigma}$; enters the Tolman SB relation \\
		$\gamma_\rho$ & Volumetric evolution exponent: $\rhoL\propto(1+z)^{\gamma_\rho}$; enters the \citet{Li2023} DDR reduction \\
		$\gamma$ & Single exponent notation adopted after the TNG result in Sec.~\ref{ssec:tng_ladder} finds $\gamma_\Sigma\simeq\gamma_\rho$ for the matched bright-galaxy sample \\
		$\Sigma_L$ & Rest-frame projected luminosity density $L/A_{\rm phys}$ \\
		$\rhoL$ & Rest-frame volumetric luminosity density $L/V$ \\
		$A_{\rm phys}$, $V$ & Rest-frame projected area and volume of the source \\
		$\Isbobs$ & Observed-frame surface brightness $F_\nu/A_{\rm ap}$ \\
		$F_\nu$ & Band-averaged total flux density of the source in a given broad filter (Sec.~\ref{ssec:obs_data}) \\
		$A_{\rm ap}$ & Observed-frame aperture area $\pi(\texttt{apopt}/2)^2$ \\
			$\alphaTol$ & Tolman redshift-scaling exponent: $\Isbobs\propto(1+z)^{\alphaTol}$ \\
			$\alphaDDR$ & DDR redshift-scaling exponent: $\DLum/\DAng^{\rm Li}\propto(1+z)^{\alphaDDR}$ \\
		$\alphaSED$ & Rest-frame SED slope: $F_\nu^{\rm rest}\propto\nu^{-\alphaSED}$ \\
		$\DLum,\DAng$ & Luminosity and angular-diameter distances \\
		$\DAng^{\rm Li}$ & Angular-diameter distance inferred via the \citet{Li2023} reduction \\
		$d_0$, $L_0$, $a$, $b$ & \citet{Li2023} size/luminosity normalisations and \\
		& joint evolution exponents (\cref{eq:Li}) \\
		$p_{\rm spec}$ & Mock spectroscopic-selection probability for a TNG galaxy \\
		\bottomrule
	\end{tabular}
\end{table}

We use the ASTRODEEP-JWST catalogue \citep{Merlin2024}, distributed as VizieR catalogue \texttt{J/A+A/691/A240}\footnote{\url{https://vizier.cds.unistra.fr/viz-bin/VizieR-3?-source=J/A+A/691/A240}}. The release contains $531\,173$ photometric objects across six extragalactic deep fields imaged with JWST ERS and Cycle-1 programmes (Abell~2744, CEERS, JADES-GN, JADES-GS+NGDEEP, PRIMER-COSMOS and PRIMER-UDS), in 16 HST/JWST broad bands. Source detection is performed with \texttt{SExtractor} \citep{SExtractor} on stacked NIRCam long-wavelength images; aperture photometry is then carried out with \texttt{a-phot} \citep{Merlin2024} on the PSF-matched single-band images, in a fixed set of nine circular aperture diameters $\{0.20, 0.28, 0.33, 0.50, 0.66, 0.70, 1.32, 2.65, 5.30\}$\,arcsec. A per-source ``optimal'' aperture (catalogue column \texttt{apopt}) is then selected by \citet{Merlin2024} from this set on the basis of the SExtractor \texttt{ISOAREA} of the source, and the corresponding flux is used to define colours; the catalogue's total-flux columns are aperture-corrected total fluxes derived from the Kron-aperture flux in the detection band. Following the prescription of \citet{Merlin2024} the per-band total flux entries \texttt{FF}$_{\rm band}$
(the VizieR column naming: e.g.\ \texttt{FF200W} is documented as the F200W total flux in $\mu$Jy, with uncertainty \texttt{e\_FF200W}) in the catalogue are computed as $f_{\rm tot, band}=c_{\rm ap,det}\,f_{\rm ap,band}$, where $f_{\rm ap,band}$ is the in-\texttt{apopt} flux and $c_{\rm ap,det}=f_{\rm tot,det}/f_{\rm ap,det}$ is the Kron-aperture correction measured on the F356W+F444W detection stack. We use these total-flux columns as $F_\nu$ throughout, i.e.\ $F_\nu\equiv f_{\rm tot,band}$ for the filter under consideration ($f_{\rm tot,det}$ enters only through the aperture correction). The subscript $\nu$ denotes a specific (per-unit-frequency) flux density, but the measurement is band-averaged: $F_\nu$ is the flux density of the relevant broad filter in $\mu$Jy, referenced to its effective frequency, not a monochromatic measurement.
We then construct the surface-brightness observable $\Isbobs=F_\nu/A_{\rm ap}$ where $A_{\rm ap}=\pi(\texttt{apopt}/2)^2$. The observable is therefore a catalogue SB proxy rather than a literal aperture-flux SB; its redshift scaling matches that of a true SB up to the extent to which the per-source \texttt{apopt} tracks the source angular size. This is the same convention as \citet{TolmanJWST}. Spectroscopic redshifts are provided where available; the remainder carry photometric redshifts from a multi-code consensus.

We adopt the same sample-definition cuts as \citet{TolmanJWST}, in order to reproduce their published measurement before extending the analysis. In their notation, these are a confirmed spectroscopic redshift $\zspec>0$, an upper-redshift cap $z\leq 14$
(applied to the catalogue best-available redshift $z$, which coincides with $\zspec$ for every object with $\zspec>0$) to remove a small number of catastrophic photometric solutions that survive the spectroscopic cut as artefacts, the catalogue photometric flag $<80$ which excludes objects flagged as image artefacts, blends, edge-of-mosaic detections or point-like sources \citep{Merlin2024}, a positive optimal-aperture radius which discards objects for which \texttt{apopt} did not return a usable aperture, and a signal-to-noise threshold of five in at least one filter%
. These cuts are deliberately minimal, simply picking out objects with reliable spectroscopic redshifts and usable photometry across the broad-band filter set. Several of these cuts---most obviously the $\zspec>0$ requirement, but also the signal-to-noise threshold---induce an effective redshift-dependent selection. Rather than being replicated cut-by-cut, this selection enters the forward model empirically: the mock-spectroscopic classifier of Sec.~\ref{ssec:obs} is trained with precisely these post-cut spectroscopic confirmations as labels, and its features include the channels through which the cuts act (the above-threshold filter count, the aperture area and a redshift proxy). 
These cuts yield $7\,059$ objects on the full filter set and $7\,056$ when restricted to the 13 filters in common with the TNG mock products (ACS F435W, F606W, F814W; WFC3 F125W, F140W, F160W; NIRCam F090W, F115W, F150W, F200W, F277W, F356W, F444W).
Adding the cut \texttt{n\_valid\_filters} $\geq 8$ yields $6\,867$ objects, within seven of the dataset size of $6\,860$ reported in \citet{TolmanJWST}.

For the distance-duality benchmark of Sec.~\ref{ssec:astrodeep_meas} we use the two ultracompact very-long-baseline interferometry samples assembled by \citet{Li2023}: the JJ2.29 compilation of \citet{JacksonJannetta2006}, as tabulated in the data file distributed with \citet{Li2023}, which contains 613 sources, and GKF5.0 \citep{GurvitsKellermannFrey1999}, distributed as VizieR catalogue \texttt{J/A+A/342/378/table1}\footnote{\url{https://vizier.cds.unistra.fr/viz-bin/VizieR-3?-source=J/A+A/342/378/table1}}, which contains 330 sources, for a combined sample of 943 objects. Each source has a tabulated angular size and a spectroscopic redshift; the two samples together span $z{=}0.0035$--$3.79$ (JJ2.29) and $z{=}0.011$--$4.72$ (GKF5.0), with ${\sim}22\%$ of each below $z=0.5$ and the bulk of the datapoints at $0.5\lesssim z\lesssim 2$.
We follow the reduction convention of \citet{Li2023} by calibrating the JJ2.29 and GKF5.0 catalogues separately (i.e. fitting the $a$, $b$ and $d_0$ parameters introduced in Sec.~\ref{ssec:astrodeep_meas}), because their angular-size definitions and observing frequencies differ. Only then do we concatenate the two samples and bin into 27 approximately equal-population redshift bins for the combined DDR slope. Luminosity distances are assigned by evaluating an external supernova Ia-calibrated distance--redshift relation at each source’s redshift, while angular diameter distances (up to a normalisation ambiguity discussed in Sec.~\ref{sec:ddt}) come from the radio sources' angular sizes.

\subsection{\texttt{IllustrisTNG-100} mock imaging}\label{ssec:tng}

\texttt{IllustrisTNG}~\citep{Marinacci2018,Naiman2018,Nelson2018,Pillepich2018,Springel2018,NelsonHLSP} is a suite of cosmological magneto-hydrodynamical simulations of galaxy formation, run with the moving-mesh code \texttt{AREPO}~\citep{Springel2010} on a \emph{Planck} 2015 background cosmology~\citep{Planck2016}. The TNG galaxy-formation model couples gravity, ideal magnetohydrodynamics, primordial and metal-line cooling, star formation with stellar evolution and chemical enrichment, galactic winds and a two-mode supermassive-black-hole feedback prescription, calibrated against the $z=0$ stellar-mass function, cosmic star-formation-rate density and galaxy gas fractions~\citep{Weinberger2017,Pillepich2018Model}. We work with the flagship \texttt{TNG100-1} run, which evolves $2\times 1820^{3}$ gas and dark-matter resolution elements in a $110.7~\rm cMpc$ box at a baryonic mass resolution of $1.4\times 10^{6}~\rm M_{\odot}$---large enough that the publicly released $137$\,arcmin$^2$ JWST/HST lightcone yields a statistically useful population of $z\lesssim 7$ galaxies at the depth probed by ASTRODEEP.

We use the lightcone field~\texttt{tng100-7-6-xyz\_137sqarcmin} from the MAST Illustris HLSP archive\footnote{\url{https://archive.stsci.edu/hlsp/illustris}}, which spans $137$\,arcmin$^2$ and provides post-processed mock images and \texttt{OutputData} catalogue tables for thirteen HST and JWST broad bands.
The mock images are noise-free and point-spread-function (PSF)-free representations of the simulated galaxy population, with SB in $\rm nJy\,pix^{-1}$. The catalogue \texttt{total\_quant} field gives the total filter flux of each subhalo, and \texttt{photrad\_kpc} defines a fixed physical aperture radius (twice the half-light radius).
Summing the image flux across the full $137$\,arcmin$^2$ lightcone and comparing band-by-band against the sum of \texttt{total\_quant} over all $7\times10^{4}$ catalogue sources, the two agree at the fractional ${\sim}10^{-5}$ level in every HST and JWST filter.
We adopt TNG100 rather than the higher-resolution TNG50~\citep{Pillepich2019}
because the TNG50 lightcone of comparable area retains only ${\sim}100$ galaxies after empirical spectroscopic selection, providing insufficient statistics for a precise slope measurement.

The SB of a TNG mock galaxy is not uniquely defined. The slope $\alphaTol$ depends sensitively on how the aperture is drawn around each object, whether and how sources are deblended and whether the aperture is held fixed in physical units or matched to the redshift-dependent ASTRODEEP \texttt{apopt} convention. In the absence of an unambiguously optimal method, we measure the slope under several physically motivated definitions and take the spread between them as a systematic uncertainty on the TNG forward-model prediction. Specifically, we define four main variants of the SB observable:%
\begin{enumerate}
	\item  Catalogue-flux SB using \texttt{total\_quant} divided by the \texttt{photrad\_kpc} aperture area;
	\item Circular-aperture image SB at the catalogue position in the \texttt{photrad\_kpc} aperture using the in-house pixel-sum aperture routine released with the code;
	\item Voronoi-deblended image SB, obtaining flux by summing pixels assigned to the nearest catalogue position by Voronoi tessellation and using the same \texttt{photrad\_kpc} area denominator;
	\item An adaptive variant, where the image flux is remeasured at the ASTRODEEP median \texttt{apopt} %
diameters ($1.32''$, $0.50''$, $0.33''$, $0.28''$ in the $z<2$, $2<z<3$, $3<z<4$, $z>4$ bins, corresponding to circular-aperture radii $0.66''$, $0.25''$, $0.165''$, $0.14''$). The Tolman SB observable uses this same redshift-dependent area as the denominator in $I_{\rm SB}=F_\nu/A_{\rm ap}$, whereas the DDR $\rho_L=L/V$ observable uses the adaptive aperture only to remeasure the flux, holding the volume fixed at the intrinsic TNG \texttt{photrad\_kpc} value (Sec.~\ref{ssec:tng_ladder}). This avoids building the imposed redshift dependence of the aperture rule into the fitted slope by construction.
\end{enumerate}
A Gaussian-weighted deblending calculation is also used separately as a diagnostic in the systematic budget.

We adopt definition~(iii), the Voronoi-deblended image SB, as fiducial: it operates on the mock images rather than the catalogue (so does not assume that pixel sums are calibrated to \texttt{total\_quant}), and it explicitly reassigns flux from regions where neighbouring objects overlap on the sky, which we find to be the dominant systematic correction in the image-aperture variants.
This variant is also the most comparable to the ASTRODEEP $F_\nu/A_{\rm ap}$ observable, in which $F_\nu$ is likewise intended as an uncontaminated total flux of the source alone---although ASTRODEEP arrives at it by aperture-correcting a narrower aperture flux rather than by deblending. (ASTRODEEP itself performs no flux-level deblending: overlapping sources are separated only at the \texttt{SExtractor} detection stage, and photometry flagged as blended is excluded by the flag cut of Sec.~\ref{ssec:obs_data} rather than corrected.)
Each of the four main variants is a well-defined
SB observable that emphasises a different aspect of the measurement chain (catalogue versus image, deblending convention, fixed versus adaptive aperture), and the agreement or disagreement between them quantifies how much of the residual difference with ASTRODEEP can be assigned to a specific photometric choice. We caution that the Voronoi step assigns each segment pixel to the nearest catalogue position rather than performing PSF-deblending of overlapping point spread functions. The recovered flux therefore depends on the segmentation threshold and on the catalogue positional accuracy, and recovers blended flux only insofar as the underlying segmentation has separated the blend in the first place.

The TNG mock-spectroscopic sample is used as input to the distance-duality test in Sec.~\ref{ssec:tng_ladder}, where we additionally compute a per-galaxy luminosity density $\rho_{L,i}=L_i/V_i$, with $L_i$ the rest-frame broad-band luminosity of the $i^{\rm th}$ galaxy (the specific band is fixed in Sec.~\ref{ssec:tng_ladder}) and $V_i=4\pi r_{{\rm phys},i}^3/3$.
For the catalogue, fixed-physical image-aperture, and Voronoi-deblended variants, $r_{\rm phys}$ is the native TNG \texttt{photrad\_kpc} aperture radius. For the adaptive-flux variant, only the flux is remeasured in the redshift-dependent ASTRODEEP-style aperture; the volume entering $\rho_L$ is still held fixed at the intrinsic \texttt{photrad\_kpc} volume, so that a fitted evolution exponent does not follow the imposed redshift dependence of the aperture rule.
The four main aperture variants listed above for the SB observable propagate one-to-one to the corresponding variants used for the DDR analysis so that the two tests share the same forward model and selection procedure. The evolution exponents the two tests constrain (described below) may however differ in general.

\section{Method}\label{sec:method}

\subsection{Evolution terms in the Tolman and DDR tests}\label{ssec:unified}

The Tolman measurement responds to projected luminosity density $\Sigma_L=L/A_{\rm phys}$, for which we write $\Sigma_L\propto(1+z)^{\gamma_\Sigma}$. The DDR reduction of \citet{Li2023} instead responds to volumetric luminosity density $\rhoL=L/V$, with $V=(4/3)\pi r^3$, for which we write $\rhoL\propto(1+z)^{\gamma_\rho}$. These exponents describe different physical constructions and must not be assumed equal.
We define the Tolman scaling as
\begin{equation}\label{eq:Isb_scaling}
\Isbobs(z) \propto (1+z)^{\alphaTol},
\end{equation}
where, under the luminosity-evolution model, the exponent is
\begin{equation}\label{eq:Isb_unified}
\alphaTol=-(3+\alphaSED)+\gamma_\Sigma.
\end{equation}
For the duality inference, applying the constant-$\rhoL$ assumption of \citet{Li2023} to a population with true evolution $\rhoL\propto(1+z)^{\gamma_\rho}$ biases the inferred angular-diameter distance multiplicatively,
\begin{equation}\label{eq:DALi_bias}
    \DAng^{\rm Li}(z) = \DAng^{\rm true}(z)\,(1+z)^{\gamma_\rho/3}.
\end{equation}
The cube root is the compact-source $L\propto d^3$ mapping adopted in the \citet{Li2023} reduction, equivalent to inverting $L=\rhoL V$ at fixed $\rhoL$.

We describe the DDR redshift scaling as
\begin{equation}\label{eq:DLDA_scaling}
	\frac{\DLum(z)}{\DAng^{\rm Li}(z)} \propto (1+z)^{\alphaDDR},
\end{equation}
where, under the \citet{Li2023} compact-source model, the exponent is
\begin{equation}\label{eq:DLDA_unified}
	\alphaDDR=2-\gamma_\rho/3.
\end{equation}
The reference values are separately $\gamma_\Sigma=0$ and $\gamma_\rho=0$ for no intrinsic evolution in the two routes, while a value of 3 moves the corresponding idealised observable to its apparently non-expanding reference.
Inferring $\gamma_\Sigma$ from Tolman SB depends on the population-averaged rest-frame SED slope $\alphaSED$ and on any aperture-area--redshift evolution introduced by the measurement rule, while inferring $\gamma_\rho$ from DDR depends on the compact-source mapping assumed by \citet{Li2023}. We therefore keep these routes explicitly separate until the TNG result motivates a shared notation.

Although it is free of $\alphaSED$, the DDR analysis has a different systematic uncertainty. The \citet{Li2023} DDR study is based on ultracompact radio sources, which are unresolved in the TNG simulation. We instead measure $\gamma_\rho$ in TNG for JWST-selected bright galaxies, not for analogues of the ultracompact-radio sources used in \citet{Li2023}, which are parsec-scale AGN structures unresolved in the simulation. Comparing TNG with the radio result therefore provides constraining power on whether these populations have similar volumetric luminosity-density evolution within the FRW framework of the simulation. (As stated above, the Tolman and DDR relations as defined here are expected to hold for any metric theory of gravity with photon number conservation, not just FRW; however, due to our direct comparison with galaxy samples from the $\Lambda$CDM TNG simulation we will refer to these fiducial relations as ``FRW''.)

\subsection{Observable calculation and mock-spectroscopic selection}\label{ssec:obs}

For each galaxy and filter, the observed-frame SB proxy is
\begin{equation}\label{eq:sb}
\Isbobs(\nu_{\rm obs}) = \frac{F_\nu(\nu_{\rm obs})}{A_{\rm ap}},
\end{equation}
the filter flux density divided by the aperture area in arcsec$^2$, with $\nu$ denoting frequency throughout. In ASTRODEEP, $F_\nu$ is taken from the catalogue \texttt{FF}$_{\rm band}$ column---i.e.\ the aperture-corrected total flux of \citet{Merlin2024} described in Sec.~\ref{ssec:obs_data}, not the in-\texttt{apopt} flux---so $\Isbobs$ is the catalogue SB proxy defined there. For a non-evolving source observed in a fixed-physical aperture, $\Isbobs\propto(1+z)^{-(3+\alphaSED)}$, where $\alphaSED$ is the rest-frame SED slope defined by $F_\nu^{\rm rest}\propto\nu^{-\alphaSED}$. The bolometric Tolman exponent $-4$ is recovered if the rest-frame specific intensity is integrated over frequency rather than measured at a single observer-frame band.
This exponent does not depend on $\alphaSED$, which is removed by integrating $F_\nu$ over $\nu$.

To estimate $\alphaSED$ we again utilise TNG100. Fitting $\log F_\nu$ against $\log\lambda_{\rm eff}$ to the 2901 mock-spectroscopic galaxies in the bands bracketing F200W gives median effective slopes $\alphaSED=0.31$ for catalogue fluxes, $0.34$ for fixed-aperture image fluxes and $0.29$ for segmentation fluxes. The corresponding 16th--84th percentile ranges are approximately $0.08$--$0.58$, $0.10$--$0.63$ and $0.08$--$0.54$, respectively. A three-band F150W/F200W/F277W fit gives indistinguishable medians. We therefore take $\alphaSED=0.3$ as our fiducial local broad-band slope for the selected bright-galaxy population, with systematic uncertainty of $\sim$0.3 that shifts the Tolman-mapped $\gamma_\Sigma$ and TNG K-corrected $\gamma_\rho$ values coherently. Directly estimating the same number from ASTRODEEP would require a full SED/noise/filter-throughput treatment, which is beyond our scope here.

For each galaxy, we obtain one $\Isbobs$ value per filter from~\cref{eq:sb}. We summarise the results across filters with mean and maximum SB observables, then fit the slope across galaxies.
We also report two further mode-style quantities. The modal mean (resp.\ modal maximum) SB is constructed by binning galaxies in $\log(1+z)$, building a histogram of their mean (resp.\ maximum) $\log\Isbobs$ values within each bin, and computing a count-weighted average over all histogram bins whose count is within $\sqrt{N_{\rm mode}}$ of the modal-bin count $N_{\rm mode}$ (rather than the modal bin alone, which would be unstable for the typical $N_{\rm mode}\lesssim 20$). The slope is then fitted against the bin-centre redshift, with each bin's uncertainty taken as the half-spread of the near-mode bin centres entering the average. This traces the typical galaxy of the population at each redshift rather than the population mean and is far less sensitive to bright-tail outliers, complementing the simple mean/max statistics. We follow \citet{TolmanJWST} in reporting all four statistics, although we take the per-galaxy mean SB as fiducial. Power-law fits of the form $\log\Isbobs=\beta+\alphaTol\log(1+z)$ are then performed by ordinary least-squares (OLS) regression.

The dominant systematic on the data side is the dependence of $\alphaTol$ on the precise sample-definition cuts; we therefore re-run the fit with each cut individually relaxed or tightened (varying the minimum number of valid filters, the per-filter signal-to-noise threshold, and so on), and report the half-range of the resulting slope values as an approximate systematic uncertainty that supplements the statistical uncertainty on the fitted slope.

We now turn from the SB observable itself to the question of which galaxies enter the slope fit.
The ASTRODEEP catalogue is an entire-population photometric catalogue, but the Tolman fit is performed only on the subset of galaxies that have a confirmed spectroscopic redshift. Which galaxies make it into that subset is determined by the spectroscopic targeting strategy of the surveys that fed redshifts back into ASTRODEEP, and reflects magnitude limits, colour preferences, instrument-specific selection, and emission-line strength criteria that we cannot model from first principles for the underlying TNG population. To compare like with like, we need to apply an analogous selection to the TNG mock galaxies before fitting any Tolman slope. Note however that we do not have spectra for the TNG mocks: we therefore approximate this selection \emph{photometrically}, asking whether each TNG mock galaxy is photometrically similar to the kind of ASTRODEEP galaxy that ended up with a confirmed spectroscopic redshift.
This makes the comparison necessarily approximate, but more than sufficient to determine whether galaxy evolution effects are expected to produce an evolution exponent approximately sufficient to reconcile the FRW expectation with the data.

To do this, we train a binary logistic-regression classifier on the ASTRODEEP catalogue. We use the default linear log-odds model of \texttt{scikit-learn} with $\ell_2$ regularisation, median imputation of missing entries and standard rescaling of inputs~\citep{Pedregosa2011}. The classifier is given an input vector of eight features: the AB magnitudes in F200W and F356W, the colours F150W$-$F356W and F200W$-$F444W, the logarithmic mean SB $\log\Isbobs$ across the available filters, the logarithmic aperture area $\log{A_{\rm ap}}$, the number of filters detected at signal-to-noise above the catalogue threshold of $5$, and \texttt{z\_proxy}, which is the catalogue $z_\text{phot}$ where available and the catalogue $z$ otherwise. \texttt{z\_proxy} is included so that the classifier can encode the well-known ASTRODEEP spectroscopic-targeting trend with redshift, but we check that removing it gives near-identical results.
The binary target is $\zspec>0$, i.e.\ whether the galaxy has a confirmed spectroscopic redshift entry in the catalogue; the classifier is therefore assessing whether a given galaxy looks photometrically like an ASTRODEEP source for which a spectroscopic redshift was successfully obtained. Spectroscopically confirmed galaxies are a small fraction of the catalogue, so we reweight the two classes during training to have equal effective populations, preventing the classifier from converging on the trivial solution ``predict no spectrum'' for every input. The model achieves a
receiver-operating-characteristic area under the curve of $0.93$, with a five-fold stratified held-out estimate of $0.933\pm 0.002$. We apply this trained classifier to the TNG mock photometry.

\begin{figure}
\centering
\includegraphics[width=\columnwidth]{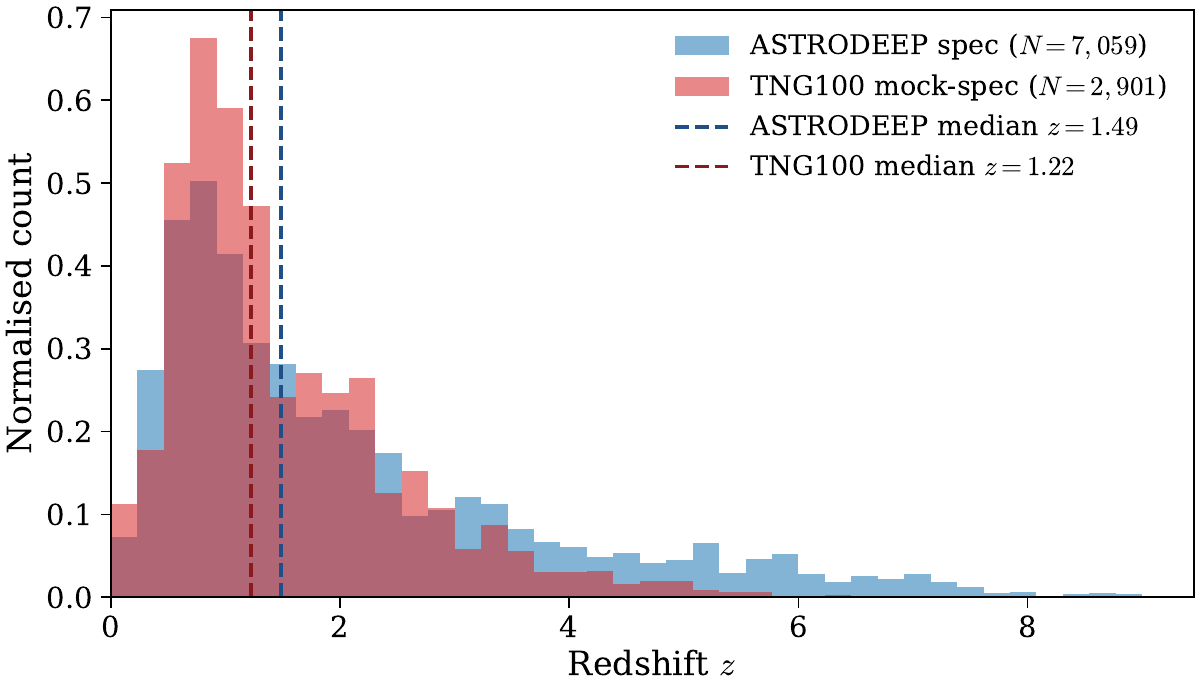}
\caption{Redshift distributions of the ASTRODEEP spectroscopic-quality sample (blue)
and the TNG100 mock-spectroscopic sample (red).
The empirical photometric classifier reproduces the bulk of the ASTRODEEP redshift distribution on the TNG mocks, but somewhat underweights the high-$z$ tail.
}
\label{fig:zdist}
\end{figure}

The output is a probability $p_{\rm spec, i}$ that the $i^\text{th}$ TNG galaxy would have been observed spectroscopically. We retain galaxies by deterministic top-$K$ selection (ranking by $p_{\rm spec, i}$ in descending order and keeping the top $K$ objects), where the count $K$ is set to the integer nearest $\sum_i p_{\rm spec,i}$. This calibrates the sample size to the integrated selection probability while keeping the highest-probability galaxies.
Of the two TNG SB pipelines defined in Sec.~\ref{ssec:tng}, deterministic selection retains $K=2\,901$ actual rows for the image-aperture pipeline (including variants ii--iv) and $K=3\,704$ actual rows for the catalogue-flux pipeline (variant i). In each case, $K$ is fixed by the rounded integrated selection probability before retaining the top-ranked objects. The catalogue-flux row uses only the MAST catalogue quantities, whereas the image-aperture rows additionally require successful pixel-level aperture extraction in the mock images.
We compare the mock and observed redshift distributions in Fig.~\ref{fig:zdist}, finding good general agreement.

\section{Results}\label{sec:results}

\label{ssec:gamma}%

\subsection{Reproducing the literature measurements}\label{ssec:astrodeep_meas}

We begin by reproducing the published Tolman and DDR analyses of \citet{TolmanJWST} and \citet{Li2023} on the data side, before turning to the TNG forward model in Sec.~\ref{ssec:tng_ladder}.

\subsubsection{Tolman test}

\begin{figure*}
\centering
\includegraphics[width=\textwidth]{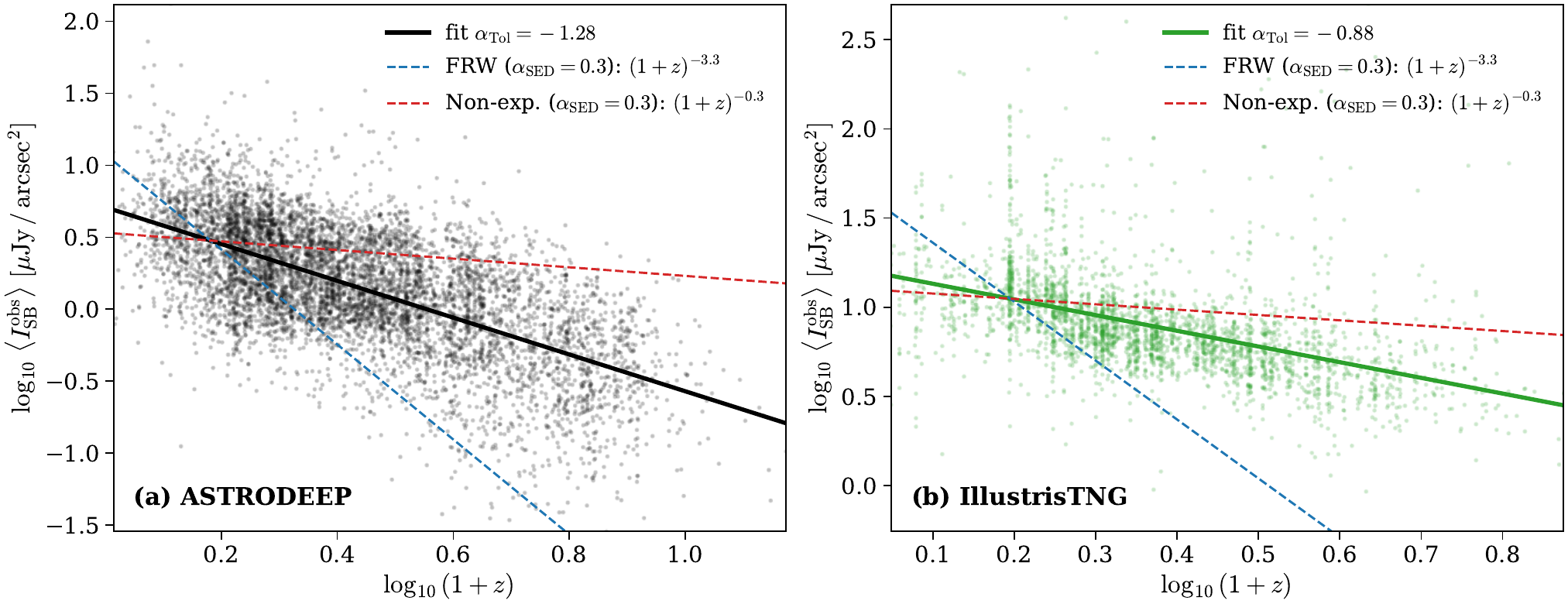}
\caption{Mean SB redshift evolution. \textbf{(a)} ASTRODEEP TNG-common-filter spectroscopic sample (black points), $N=7\,056$; fiducial slope $\alphaTolMean^{\rm ASTRODEEP}=-1.28\pm 0.05$. \textbf{(b)} TNG100 Voronoi-deblended mock-spectroscopic sample (green points), $N=2\,901$; fitted slope $\alphaTolMean^{\rm TNG}=-0.879\pm 0.030$. In each panel, the solid line is the best-fit power law, while blue-dashed shows the $\alphaSED=0.3$ broad-band FRW scaling $(1+z)^{-3.3}$ and red-dashed the corresponding non-expanding scaling $(1+z)^{-0.3}$. The measured ASTRODEEP and TNG slopes lie between these references.}
\label{fig:meansb}
\end{figure*}

Our result for the JWST Tolman analysis of \citet{TolmanJWST} on the spectroscopic-quality sample, restricting to the 13 TNG-common filters, is shown in Fig.~\ref{fig:meansb}(a). The OLS fit gives $\alphaTolMean^{\rm ASTRODEEP}=-1.279\pm 0.018$. Combining filters using the ``max'' rather than ``mean'' method of Sec.~\ref{ssec:obs} yields $\alphaTolMax^{\rm ASTRODEEP}=-1.130\pm 0.020$,
and the corresponding modal mean and modal maximum slopes are $-1.039\pm 0.134$ and $-1.101\pm 0.095$ respectively. These are consistent with the results of \citet{TolmanJWST}, who quote $-1.26\pm 0.06$ for the mean SB and $-1.10\pm 0.06$ for the maximum SB on essentially the same sample of ${\sim}6\,860$ galaxies.
This gives us confidence that we have correctly reproduced the published ASTRODEEP measurement and allows us to investigate systematics on it before turning to the TNG forward model.
This reproduces the measured ASTRODEEP slopes, rather than the cosmological reference curves overplotted in fig.~2 of \citet{TolmanJWST}. Since those data are flux-density quantities in broad filters, the appropriate no-evolution references at our fiducial $\alphaSED=0.3$ are $\alphaTol=-3.3$ for FRW and $\alphaTol=-0.3$ for a non-expanding scaling, rather than the bolometric $-4$ and $-1$ curves they show. The data therefore lie between these limiting tracks, as visible in Fig.~\ref{fig:meansb}; their corresponding projected-evolution value is $\gamma_\Sigma=2.02\pm0.05$.
Also shown in Fig.~\ref{fig:meansb} are lines corresponding to the broad-band FRW (blue) and non-expanding (red)
expectations assuming no source evolution.

Relaxing or tightening the documented sample-definition cuts---\texttt{n\_valid\_filters} $\geq 8$, or any-filter S/N $\geq 10$---shifts $\alphaTolMean$ by $\Delta\alphaTol\in\{-0.002,+0.089\}$, giving a cut-to-cut systematic of $0.045$ on the ASTRODEEP slope. We therefore quote the fiducial ASTRODEEP mean-SB uncertainty as $\sqrt{0.018^2+0.045^2}\simeq0.05$, combining the OLS statistical error and the cut systematic in quadrature.

\subsubsection{Distance-duality test}\label{sec:ddt}

\citet{Li2023} parametrises the luminosity evolution of the ultracompact-radio source population as
\begin{equation}\label{eq:Li}
	L(z)=L_0\left[(1+z)^a\,\ln(1+z)^b\right]^3,
\end{equation}
and the linear-size relation as
\begin{equation}
	l(z)=d_0\left[(1+z)^a\,\ln(1+z)^b\right].
\end{equation}
These share exponents $a$ and $b$, while the latter introduces $d_0$ as the rest-frame standard ruler of physical size.
Our joint inference of $a$, $b$ and $d_0$ on JJ2.29 and GKF5.0 agrees with \citet{Li2023} to within their stated uncertainties (Table~\ref{tab:li_repro}).
\begin{table}
\centering
\footnotesize
\setlength{\tabcolsep}{3pt}
\caption{Reproduction of the \citet{Li2023} size--luminosity calibration on the JJ2.29 and GKF5.0 ultracompact-radio samples. Each cell shows our value (left) and the published value (right).}
\label{tab:li_repro}
\begin{tabular}{lll}
\toprule
Parameter & JJ2.29 & GKF5.0 \\
\midrule
$a$           & $0.40\pm 0.05$ \,/\, $0.39\pm 0.06$ & $0.016\pm 0.083$ \,/\, $0.02\pm 0.08$ \\
$b$           & $0.604\pm 0.020$ \,/\, $0.60\pm 0.02$ & $0.718\pm 0.036$ \,/\, $0.72\pm 0.04$ \\
$d_0$ [pc] & $21.5\pm 0.4$ \,/\, $21.6\pm 0.7$ & $66.6\pm 11$ \,/\, $63.7\pm 4.4$ \\
\bottomrule
\end{tabular}
\end{table}
The binned medians of $\DLum/\DAng^{\rm Li}(z)$ are shown as the black points in Fig.~\ref{fig:li_bridge}(a), recovering the claim that the data lie \emph{parallel} to the linear non-expanding track $\DLum/\DAng=(1+z)$ rather than the FRW scaling $\propto(1+z)^2$.

It is important to realise that the absolute normalisation of this relation is not fixed by the angular-size and redshift data alone. In the \citet{Li2023} reduction it is carried by the fitted rest-frame size scale $d_0$; the luminosity normalisation $L_0$ is degenerate with the assumed source luminosity density and cancels from the plotted $\DLum/\DAng^{\rm Li}$ ratio once the size--luminosity calibration has been imposed. The $d_0$ values in Table~\ref{tab:li_repro} reproduce the \citet{Li2023} convention, under which the data---as in Fig.~\ref{fig:li_bridge}(a)---is normalised to the non-expanding line. Panel~(b) instead uses the FRW normalisation, which rescales every per-source $\DLum/\DAng^{\rm Li}$ ratio upward by $d_0^{\rm non}/d_0^{\rm exp}\simeq 2$. Panel~(c) circumvents this issue by normalising the lowest data and TNG points to 1. The fact that, as in \citet{Li2023}, the data appear on the non-expanding line in panel~(a) is therefore a feature of the imposed $d_0$ convention, not of the data alone.
However the \emph{slope} of $\DLum/\DAng^{\rm Li}$ with redshift, $\alphaDDR$, is insensitive to this issue.
Fitting a power law to the 27 binned medians gives $\alphaDDR=1.17\pm0.09$, which through~\cref{eq:DLDA_unified} corresponds to $\gamma_\rho=2.50\pm0.26$ (Table~\ref{tab:gamma_aperture}, bottom row). If instead the two raw catalogues are concatenated before fitting one shared $a$, $b$ and $d_0$, the result is $\gamma_\rho=2.06\pm0.18$; we treat this as a diagnostic rather than fiducial because it forces a common size calibration on two separate samples, which \citet{Li2023} cautions against.

\subsection{TNG forward models}\label{ssec:tng_ladder}

We first ask whether the two evolution terms that enter the tests separately in Sec.~\ref{ssec:unified} are in fact related in the simulation. For the matched fixed-physical-aperture TNG sample, the projected luminosity-density fit gives $\gamma_\Sigma=2.462 \pm 0.029$, while the volumetric luminosity-density fit gives $\gamma_\rho=2.457 \pm 0.042$ (uncertainties from the OLS fit). Their difference is statistically negligible. Thus, although projected and volumetric evolution need not agree \textit{a priori}, the state-of-the-art $\Lambda$CDM forward model finds that they do so for the bright-galaxy population relevant here. 
This indicates that the sample-averaged physical radius of the TNG galaxies has essentially no evolution, with the luminosity density evolution driven entirely by $L$, and may be because intrinsic size evolution (higher-$z$ galaxies being smaller) is compensated by the selection preferentially catching intrinsically larger/brighter objects at higher $z$.
This empirical identification of $\gamma_\Sigma$ and $\gamma_\rho$ enables us to denote both by a single exponent $\gamma$ henceforth---a crucial step in the unification of the Tolman and distance-duality tests.

The TNG Tolman slopes, i.e. variants of $\alphaTol$, from the four main aperture conventions of Sec.~\ref{ssec:tng} are summarised in Table~\ref{tab:tng_ladder} and compared graphically in Fig.~\ref{fig:fourway}. Each of these variants is a self-consistent SB measurement that emphasises a different aspect of the photometric chain (catalogue versus image-derived flux, circular versus deblended versus adaptive aperture). Their spread gives an approximate envelope to the TNG prediction by quantifying systematic uncertainty in the ASTRODEEP-matched forward model at fixed underlying galaxy population. The Voronoi-deblended TNG measurement, which reassigns flux from blended segments to nearest catalogue positions, is our headline TNG estimate
because it is the variant that most closely tracks what the ASTRODEEP \texttt{apopt} aperture is designed to do: it operates on the mock images rather than catalogue scalars (so it does not implicitly assume that pixel sums are calibrated to \texttt{total\_quant}), and it explicitly reassigns flux from regions where neighbouring objects overlap on the sky, which we find to be the dominant correction at low redshift where ASTRODEEP apertures are large. This produces $\alphaTolMean^{\rm TNG}=-0.879\pm 0.030$, far shallower than the broad-band FRW reference of $-3.3$ at $\alphaSED=0.3$, and $\sim\!0.4$ less steep than the ASTRODEEP value. The TNG modal-maximum slope, $-1.009\pm 0.052$, is consistent with ASTRODEEP within the joint statistical uncertainty. The full data and best-fit power law are shown for the Voronoi-deblended aperture variant in~Fig.~\ref{fig:meansb}(b).

\begin{table*}
\centering
\caption{
		Observed and simulated Tolman exponents $\alphaTol$ for the SB--redshift relation. Quoted uncertainties are OLS standard errors on the power-law fit. This is supplemented by the systematic envelope ($\sim$0.045 in $\alphaTol$) on the data side and the aperture-convention spread ($\sim$0.20) on the simulation side. $N$ is the sample size in each case.
	}
\label{tab:tng_ladder}
\begin{tabular}{lrrrrr}
\toprule
Measurement & $N$ & $\alphaTolMean$ & $\alphaTolMax$ & $\alphaTolModalMean$ & $\alphaTolModalMax$ \\
\midrule
ASTRODEEP TNG-common              & $7\,056$ & $-1.279\pm 0.018$ & $-1.130\pm 0.020$ & $-1.039\pm 0.134$ & $-1.101\pm 0.095$ \\
TNG catalogue-flux                & $3\,704$ & $-1.015\pm 0.019$ & $-1.003\pm 0.018$ & $-0.984\pm 0.044$ & $-0.841\pm 0.112$ \\
TNG circular image aperture       & $2\,901$ & $-0.718\pm 0.030$ & $-0.703\pm 0.029$ & $-0.808\pm 0.058$ & $-0.881\pm 0.062$ \\
TNG Voronoi-deblended (headline)  & $2\,901$ & $-0.879\pm 0.030$ & $-0.866\pm 0.029$ & $-0.799\pm 0.082$ & $-1.009\pm 0.052$ \\
TNG ASTRODEEP-like aperture       & $2\,901$ & $+0.695\pm 0.046$ & $+0.717\pm 0.047$ & $+2.317\pm 0.442$ & $+1.640\pm 0.299$ \\
\bottomrule
\end{tabular}
\end{table*}

\begin{figure*}
	\centering
	\includegraphics[width=\textwidth]{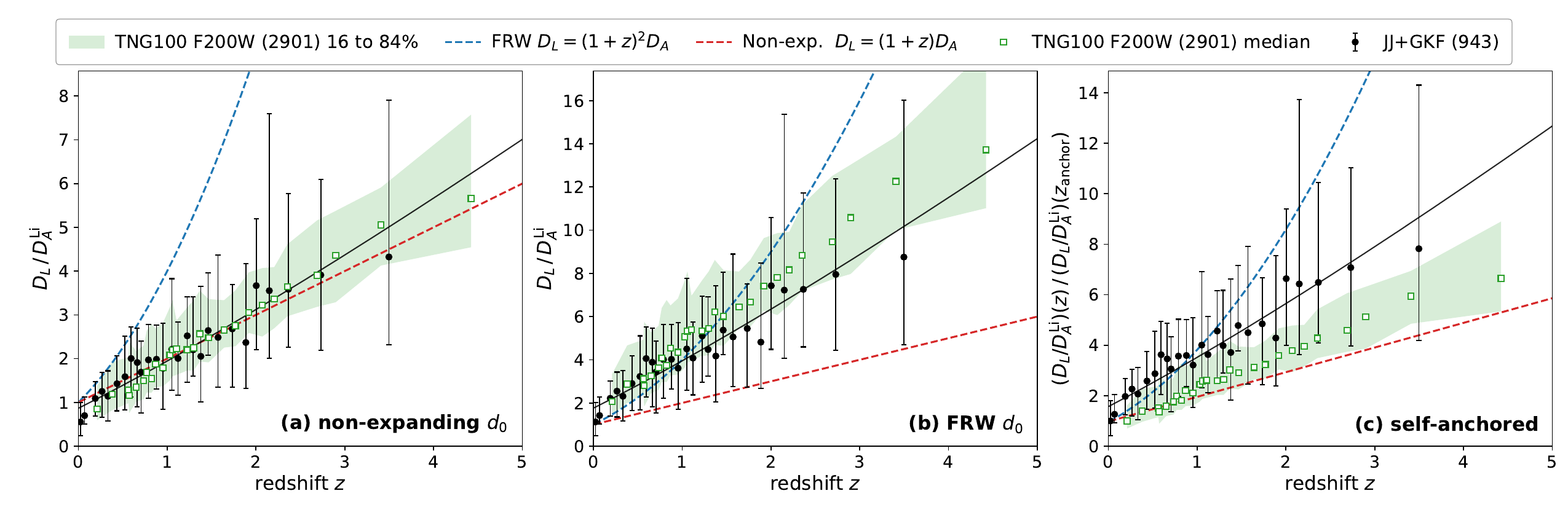}
	\caption{Three views of the same binned DDR ratio $\DLum/\DAng^{\rm Li}(z)$. Black circles: combined \citet{JacksonJannetta2006} and \citet{GurvitsKellermannFrey1999} ultracompact-radio sources reduced with the \citet{Li2023} estimator (Sec.~\ref{ssec:astrodeep_meas}); error bars give the 16th--84th percentile range of the per-source $\DLum/\DAng^{\rm Li}$ values in each redshift bin. Green squares show the TNG100 mock-spectroscopic sample reduced with the same estimator, with the band showing the corresponding 16--84 per-cent range. The black curves show the weighted power-law fit to the binned medians. The blue dashed line is the FRW prediction $\DLum=(1+z)^{2}\DAng$ and the red dashed line the non-expanding $\DLum=(1+z)\DAng$. \textbf{(a)} The \citet{Li2023} convention, in which the rest-frame size standard $d_0$ is fixed by requiring the median source to sit on the non-expanding line. \textbf{(b)} The same data and TNG sample with the FRW choice of $d_0$ instead. \textbf{(c)} Each curve divided by its own lowest-$z$ bin median, to avoid any cosmological assumption. The power-law slope of the black line is the same in each panel; only the normalisation differs.}
\label{fig:li_bridge}
\end{figure*}

\begin{figure*}
\centering
\includegraphics[width=\textwidth]{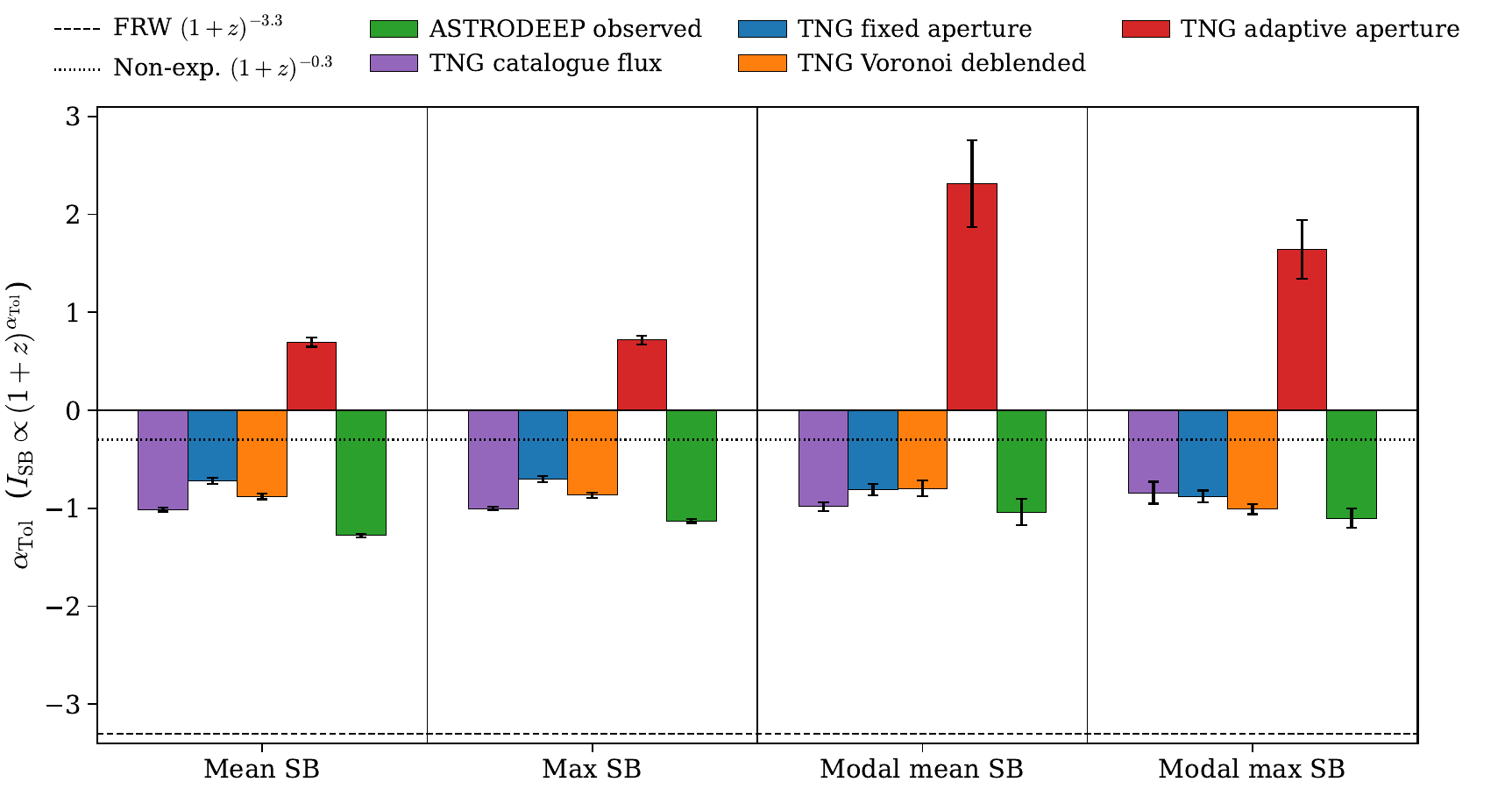}
\caption{Slope comparison for the mean, maximum, modal-mean and modal-max Tolman exponents $\alphaTol$ across the TNG measurement variants, and the ASTRODEEP reference. The errorbars show the OLS statistical uncertainties.}
\label{fig:fourway}
\end{figure*}

We next quantify the aperture dependence of the volumetric route that enters the Li-style DDR mapping.
We take the rest-frame F200W luminosity of a source to be $L_\nu=4\pi\DLum^2 S_\nu/(1+z)^{1-\alphaSED}$ (the standard K-correction for a power-law SED $F_\nu\propto\nu^{-\alphaSED}$), with $S_\nu$ the observed F200W flux. Fitting $\rhoL\propto(1+z)^\gamma$ to $L_\nu$ divided by the intrinsic \texttt{photrad\_kpc} volume gives the values in Table~\ref{tab:gamma_aperture} for the fiducial mock-ASTRODEEP selection. We show the statistical uncertainty on the OLS fits, and check that bootstrap resampling gives consistent errors to within $\sim15\%$.
In the first three variants this is the same physical aperture used to define the catalogue or image flux. In the adaptive-flux variant, the flux is measured in the redshift-dependent ASTRODEEP-style aperture, but the volume is held fixed at the intrinsic TNG \texttt{photrad\_kpc} value. This keeps $\gamma$ tied to luminosity-density evolution of the selected simulated galaxies rather than to the imposed redshift dependence of the aperture radius.

The contrast with the adaptive row of Table~\ref{tab:tng_ladder} is instructive. For any SB measurement, writing $F_\nu\propto(1+z)^{\alpha_{F_\nu}}$ and $A_{\rm ap}\propto(1+z)^{\alpha_{A_{\rm ap}}}$ gives $\alphaTol=\alpha_{F_\nu}-\alpha_{A_{\rm ap}}$. The median-aperture rule contracts $A_{\rm ap}$ by a factor of $\sim20$ between $z<2$ and $z>4$ (giving $\alpha_{A_{\rm ap}}\simeq-2.8$), which alone adds $\simeq+3$ to $\alphaTol$ and flips its sign relative to a fixed-physical aperture. It is therefore an aperture stress test rather than a check of the matched-aperture equality above. In the DDR-side calculation its fixed-volume value, $\gamma=2.12$, reflects only the residual difference in flux capture.

The four aperture variants are centred around $\gamma=2.23$, with a half-range spread of $0.20$. The TNG K-correction uses the effective local broad-band SED slope estimated in Sec.~\ref{ssec:obs}. Shifting the assumed SED slope by $\Delta\alphaSED$ shifts every TNG $\gamma$ value by the same $\Delta\alphaSED$.
Through~\cref{eq:DLDA_unified} this predicts %
$\alphaDDR=1.26\pm 0.07$, visually close to the empirical radio-source track (Fig.~\ref{fig:li_bridge}, green band and squares) and in good agreement with the $\gamma=2.50\pm0.26$ inferred from the radio data.
The slight discrepancy may be informative: at face value, the higher radio value (by $\Delta\gamma=0.27\pm0.33$ if including the half-width of the aperture variants in the uncertainty) suggests the compact-radio population evolves relative to the bright-galaxy population as $\rho_L^{\rm radio}/\rho_L^{\rm gal}\propto(1+z)^{0.27\pm0.33}$, e.g.\ by a factor $\sim1.3$ by $z=2$.
This completes the set of results needed to fully understand Fig.~\ref{fig:gamma_schematic}, including the TNG aperture envelope ($\gamma=2.23\pm0.20$), ASTRODEEP measurement ($\gamma=2.02\pm0.05$) and DDR measurement ($\gamma=2.50\pm0.26$) projected from $\alphaTol$ and $\alphaDDR$ onto the common $\gamma$ parameter.

\begin{figure}
\centering
\includegraphics[width=\columnwidth]{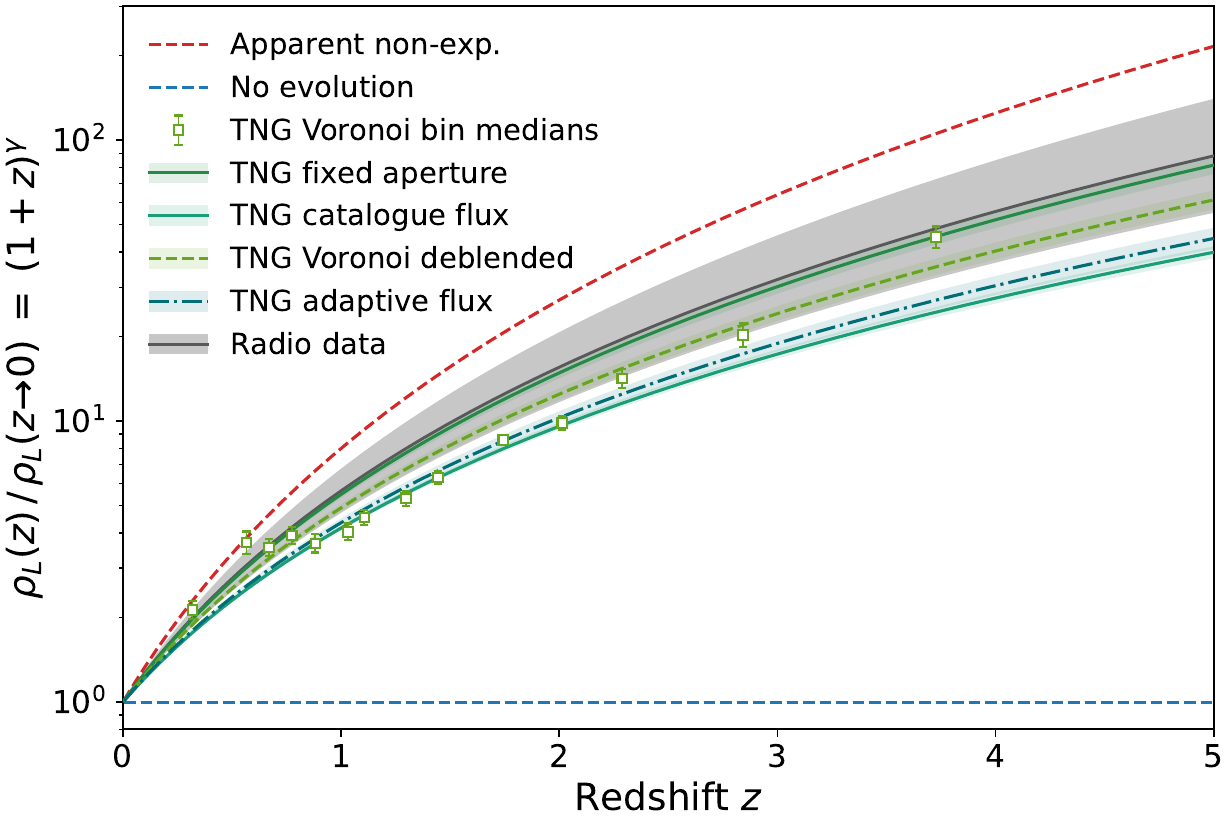}
\caption{The DDR comparison of Fig.~\ref{fig:li_bridge} reprojected onto the $\rho_L=L/V$ axis, which is physical in the simulation. The four TNG aperture variants of Table~\ref{tab:gamma_aperture} are shown as $(1+z)^\gamma$ curves with their OLS uncertainty bands; open squares give the fiducial Voronoi-deblended variant's binned medians (the other variants are similar). The grey band shows the radio-data slope $\gamma=2.50\pm0.26$. The blue dashed line indicates no evolution, and the red dashed line the evolution required to produce apparently non-expanding DDR behaviour when assuming no evolution.
}
\label{fig:rho_L}
\end{figure}

An alternative, $d_0$-free presentation of the DDR comparison is shown in Fig.~\ref{fig:rho_L}. There the $y$-axis is the population-averaged $\rho_L(z)/\rho_L(z\!\to\!0)=(1+z)^\gamma$, measured directly on TNG from the per-galaxy luminosity and aperture volume. The four TNG aperture variants appear as power-law curves with their OLS uncertainty bands, the fiducial Voronoi-deblended variant's binned medians as open squares, and the radio-data value $\gamma=2.50\pm0.26$ as a grey band. The FRW DDR prediction ($\gamma=0$) and the strict non-expanding requirement ($\gamma=3$) bracket the figure.
This plot also illustrates (alongside the normalisation-dependence of the results in Fig.~\ref{fig:li_bridge}) that the data is not ``surprisingly aligned'' with a unique non-expanding endpoint: the preferred slope is between the FRW and non-expanding limits, and the DDR normalisation cannot be established from the data.

We check explicitly that reweighting the TNG sample in $\log(1+z)$ to match exactly the ASTRODEEP redshift distribution shifts $\alphaTol$ (or equivalently its mapped $\gamma$) by only $\sim0.02$. We also test the robustness of the classifier by dropping SB-related features; this shifts the Tolman slope by only $\Delta\alphaTol\simeq-0.01$.
Adopting an $L\propto d^2$ disc mapping instead of the compact-sphere cube root gives $\gamma=3.39\pm0.11$ for the DDR data, in substantially stronger tension with TNG. The OLS standard error on $\gamma$ within each TNG variant is $0.03$--$0.05$, consistent with bootstrap resampling of the $(z,\rho_L)$ pairs to within $\sim15\%$.

To test the importance of the spectroscopic selection itself, we recompute $\gamma$ on fully random TNG subsamples of the same size $K=2\,901$. Repeated over $30$ random seeds, this yields $\gamma=3.32\pm0.09$ at $\alphaSED=0.3$. A direct calculation on the full TNG image-aperture-success pool gives $\gamma\simeq3.29$, while restricting to $z<4$ gives $\gamma\simeq3.44$. The empirical spectroscopic selection therefore moderates luminosity-density evolution relative to the full TNG population.
Importantly however, the detailed ASTRODEEP-trained classifier is not solely responsible for this shift: a generic bright-galaxy selection retaining the $K=2\,901$ brightest F200W objects gives $\gamma\simeq2.45$, close to the mock-spectroscopic value. The DDR comparison should therefore be read as a radio-AGN versus bright-galaxy luminosity-density comparison, rather than as a peculiarity of the mock-JWST spectroscopic classifier.

Although not important for the main thrust of our analysis, which concerns only the redshift slope, we note that the absolute normalisation of $\Isbobs$ also differs between TNG and ASTRODEEP, as is readily visible in Fig.~\ref{fig:meansb}. At $z<0.5$ the median F200W SB of the fiducial Voronoi-deblended TNG sample is a factor of $\sim 3$ above that of the ASTRODEEP TNG-common sample, with the ratio growing by $\sim 0.3$\,dex over $0<z<7$. The TNG and ASTRODEEP apertures at $z<0.5$ are nearly the same size---the TNG image aperture has median radius $\sim 0.51''$ and the ASTRODEEP \texttt{apopt} median radius is $0.66''$ (half the median \texttt{apopt} diameter of $1.32''$)---so the discrepancy derives almost entirely from intrinsic brightness.
This is consistent with the documented tendency of TNG to overproduce stellar mass (and hence rest-frame brightness) at the bright end at low redshift \citep{Pillepich2018,Pillepich2019,Donnari2019,Genel2018}. We note that at high redshift JWST has uncovered an apparent \emph{excess} of bright sources relative to standard $\Lambda$CDM-based galaxy-formation models %
\citep{Lovell2023,Wilkins2023,Yung2024}, so the absolute-normalisation tensions at low and high redshift appear not to have a common origin. %
\citet{Lu2025} have shown that $\Lambda$CDM-based semi-analytic models reproduce the high-$z$ UV luminosity functions from JWST to $z\simeq 10$ without invoking new physics, while the low-$z$ TNG offset at the bright end reflects the stellar-feedback calibration of TNG itself. This implies that the absolute-normalisation offset is a property of the specific simulation setup rather than of $\Lambda$CDM galaxy formation in general.

\begin{table}
\centering
\begin{tabular}{lccc}
\toprule
Variant & $N$ & $\gamma$ & $\langle r_{\rm ap}\rangle$ \\
\midrule
TNG catalogue flux + $2r_{\rm half}$ aperture   & $3\,703$ & $2.06\pm 0.03$         & $0.47$ \\
TNG fixed-phys.\ image aperture                  & $2\,901$ & $2.46\pm 0.04$         & $0.51$ \\
TNG Voronoi-deblended                            & $2\,901$ & $2.30\pm 0.04$         & $0.51$ \\
TNG adaptive flux                                & $2\,901$ & $2.12\pm 0.05$         & $0.66$ \\
\midrule
	Data Tolman (mapped $\alphaTol$) & $7\,056$ & $2.02\pm 0.05$ &---\\
Data DDR (mapped $\alphaDDR$) & $943$    & $2.50\pm 0.26$ &---\\
\bottomrule
\end{tabular}
\caption{Effective luminosity-density evolution exponent $\gamma$ from the three routes defined in Table~\ref{tab:notation}. \emph{Top block:} TNG100 values fitted from $\rhoL\propto(1+z)^\gamma$ across four aperture conventions. The catalogue-flux row uses the native TNG \texttt{photrad\_kpc} radius, i.e. twice the stellar half-light radius. The TNG and Tolman-data rows assume $\alphaSED=0.3$.
The luminosity $L$ uses the flux from each aperture variant, while the volume $V$ is the intrinsic \texttt{photrad\_kpc} volume; this differs from the flux aperture only for the adaptive-flux variant. $\langle r_{\rm ap}\rangle$ is the population-median flux-aperture radius in arcsec. \emph{Bottom block:} the ASTRODEEP entry is the effective projected exponent obtained by mapping its measured $\alphaTol$, while the DDR entry is the effective volumetric exponent obtained by mapping the radio-source $\alphaDDR$.
	The TNG and DDR uncertainties are OLS standard errors on their power-law fits; the Data Tolman entry additionally includes its systematic cut-to-cut uncertainty.}
\label{tab:gamma_aperture}
\end{table}

\label{ssec:syst}%

\section{Discussion}\label{sec:discussion}

We have shown that three routes to the exponent of luminosity density evolution with redshift are in broad agreement: TNG gives $\gamma=2.23\pm0.20$, the ASTRODEEP Tolman slope gives $\gamma=2.02\pm0.05$ (both at $\alphaSED=0.3$), and the ultracompact-radio DDR data give $\gamma=2.50\pm0.26$. While the latter two technically involve different exponents---the evolution of projected or volumetric luminosity density, respectively---these are near-identical in TNG, affording a unified explanation of the two observables.
The upshot is that no modification of standard galaxy formation in $\Lambda$CDM (and \textit{a fortiori} of FRW geometry) is required to explain either of the claimed Tolman or distance-duality anomalies.

It may seem at first that the relevant target should be $\gamma=3$, since this moves each idealised observable from its no-evolution FRW reference to its corresponding non-expanding curve, as claimed in \citet{Li2023} and \citet{TolmanJWST}. However, the data do not require this. For Tolman, the broad-band no-evolution FRW reference is $\alphaTol=-3.3$ at $\alphaSED=0.3$, and the non-expanding reference is $\alphaTol=-0.3$. The observed ASTRODEEP mean-SB slope, $\alphaTol=-1.28\pm0.05$, lies between these limits. For DDR, \cref{eq:DLDA_unified} changes the FRW exponent to $\alphaDDR=2-\gamma/3$; the normalisation-free radio fit gives $\alphaDDR=1.17\pm0.09$, again inconsistent with the non-expanding expectation $\alphaDDR=1$. The visually striking fit of the data to the non-expanding line presented in~\citet{Li2023} (and reproduced in Fig.~\ref{fig:li_bridge}(a)) rests on an arbitrary normalisation convention.

We therefore see that the shallow JWST Tolman slope \citep{TolmanJWST} and apparently non-expanding distance-duality result \citep{Li2023} should not be read as two independent anomalies requiring a coordinated modification of FRW geometry. Rather, both are determined by intrinsic population evolution. The qualitative joint sensitivity of SB and angular-size tests to source evolution has been pointed out before \citep{LopezCorredoira2015,Lerner2018}, with \citet{KhedekarChakraborti2011} proposing a 21\,cm mass--size substitute observable designed to be evolution-immune. We supply a controlled forward-model comparison of these effects in a $\Lambda$CDM simulation.
The TNG mock therefore shows that standard galaxy-formation physics can produce luminosity-density evolution of the magnitude relevant for the DDR comparison.
The small discrepancy in the DDR exponent $\alphaDDR$ carries physical information within $\Lambda$CDM: the offset between the TNG and radio values, $\Delta\gamma=0.27\pm0.33$, suggests that the unmodelled compact-radio population may evolve slightly more strongly in luminosity density than the modelled bright galaxies. Equivalently, $\rho_L^{\rm radio}/\rho_L^{\rm gal}\propto(1+z)^{0.27\pm0.33}$, a factor of $\sim1.3$ where the radio sample has most of its weight, although consistent with identical evolution within $1\sigma$.

Besides the aperture convention, six systematics could in principle move the TNG mean-SB slope:
\begin{enumerate}
	\item \emph{Redshift distribution:} The TNG mock-spectroscopic sample has a slightly lower-redshift distribution than the ASTRODEEP spectroscopic sample (Fig.~\ref{fig:zdist}). Reweighting the TNG sample to match the ASTRODEEP $z$-distribution exactly shifts the Tolman exponent by $\Delta\alphaTol\simeq+0.02$.
			\item \emph{Compactness selection bias:} Spectroscopically selected TNG galaxies could be systematically more or less extended than spectroscopically selected ASTRODEEP galaxies at fixed $z$, which would affect the SB values. We investigate this using $\log A_{\rm ap}$, the logarithm of the circular aperture area entering $F_\nu/A_{\rm ap}$, as a compactness proxy, since compact objects have smaller $A_{\rm ap}$ at fixed redshift. Splitting the TNG sample at the median residual around the population $\log A_{\rm ap}$--$z$ trend and refitting on each half gives a half-range shift of $\Delta\alphaTol\simeq 0.06$.
			\item \emph{ASTRODEEP sample-definition:} The ASTRODEEP Tolman exponent itself is sensitive to the precise sample-definition cuts. Tightening or loosening the documented cuts (Sec.~\ref{ssec:astrodeep_meas}) shifts $\alphaTolMean^{\rm ASTRODEEP}$ by $\Delta\alphaTol\in\{-0.002,+0.089\}$, indicating a total uncertainty contribution $\lesssim0.1$ (we use the half-range, $0.045$, in the fiducial total errorbar).
		\item \emph{Deblending algorithm:} Switching from our headline Voronoi deblending to a Gaussian-weighted variant (which spreads pixel flux smoothly across overlapping segments rather than reassigning it to nearest catalogue positions) shifts $\alphaTolMean$ by $\sim 0.15$---the largest single component of the budget and the dominant uncertainty on the TNG image-aperture pipeline.
		\item \emph{PSF and symmetric sky noise:} Convolving the TNG F200W image with an approximation of the JWST NIRCam PSF ($\sigma=1$\,pix) and adding Gaussian sky noise at the per-pixel depth of the deepest ASTRODEEP fields ($\sigma_{\rm pix}=0.1$\,nJy/pix) shifts $\alphaTolMean$ by $|\Delta\alphaTol|<0.005$ in either aperture convention, demonstrating that the sign reversal of the adaptive-aperture run is not driven by PSF or symmetric sky noise.
	\item \emph{Cosmic variance:} The principal TNG result is taken from a single $137$\,arcmin$^2$ line of sight (LOS; \texttt{tng100-7-6-xyz}). Repeating the fixed-physical image-aperture pipeline
    on the alternative LOS-axis-rotation lightcone \texttt{tng100-7-6-yxz} (the only other TNG100 137\,arcmin$^2$ field on the MAST archive, sharing the underlying matter cube but integrated along a different LOS) shifts $\gamma$ by only $\sim-0.05$.
\end{enumerate}
Combining these in quadrature gives $\sigma_{\alpha_{\rm Tol}}\simeq 0.19$ on the TNG--ASTRODEEP mean-SB Tolman-exponent gap, against an observed gap of $\Delta\alphaTol\simeq 0.40$ between the headline Voronoi-deblended TNG variant and the ASTRODEEP TNG-common reference. The other Tolman summary statistics in Table~\ref{tab:tng_ladder} (max SB, modal mean SB, modal max SB) have smaller gaps, with the modal-maximum Voronoi-deblended exponent already consistent with ASTRODEEP within statistical uncertainties.

Four limitations motivate clear next steps. First, the TNG mock images are noise-free and PSF-free; we have shown that PSF and symmetric Gaussian noise alone do not flip the SB slope sign, but asymmetric image-chain effects---positive sky-background residuals, source-extraction-defined aperture boundaries, detection-driven thresholding---might.
A matched PSF, sky-background and \texttt{SExtractor} reproduction is the natural next step, which could be expected to close both the residual Tolman slope gap and the absolute-normalisation offset \citep{Snyder2022}. Second, the
forward-modelling step is restricted to a single galaxy-formation prescription: cross-simulation validation against the wider-area TNG mock survey of \citet{Snyder2022}, the high-$z$ BlueTides mock-image catalogue of \citet{Marshall2022} (which extends to $z=7$--$12$ where TNG100 is statistically thin), and the Santa Cruz semi-analytic JWST lightcones of \citet{Yung2022} would test how strongly the SB slope and $\gamma$ depend on the underlying galaxy-formation model and extend the redshift baseline of the test.
Third, our empirical mock-spectroscopic selection is based on photometric features and cannot reproduce the non-photometric components of real spectroscopic targeting (emission-line strength, redshift desert, instrument-specific gratings). These have been explored at the population level \citep{Whitney2020,Fagioli2018,Fagioli2020} and are not expected to be a significant systematic, but have not been propagated quantitatively here.
Finally, making the DDR comparison fully quantitative requires either a targeted forward model of the compact-radio AGN population itself, or an alternative source population whose luminosity evolution can be independently calibrated. The latter route is illustrated by classical Tolman analyses of early-type galaxies, which used locally calibrated surface-brightness--radius relations at Petrosian metric radii and explicitly corrected for passive luminosity evolution \citep{LubinSandage2001a,LubinSandage2001b,LubinSandage2001c,LubinSandage2001d,Pahre1996}. This would reduce the source-evolution degeneracy and yield a cleaner cosmological test.

\section{Conclusion}\label{sec:conclusion}

The Tolman surface brightness and distance-duality tests are foundational in cosmology: they probe whether the Universe is expanding at all, without any need for dynamical assumptions. Both tests have recently been claimed to deliver signals indicating a breakdown in FRW cosmology: a Tolman slope much flatter than the geometric prediction and a distance-duality ratio lying on the non-expanding track.

Our main contribution has been twofold. First, we showed that the projected and volumetric luminosity-density trends of the matched TNG bright-galaxy sample are described, to excellent accuracy, by one effective evolution exponent $\gamma$, which enters the Tolman and Li-style DDR relations with different powers.
Second, and more substantively, we forward-modelled the Tolman SB measurement and fitted $\gamma$ in the \texttt{IllustrisTNG-100} simulation, including an empirical mock-spectroscopic selection trained on ASTRODEEP-JWST.

Our principal finding is that ordinary $\Lambda$CDM galaxy evolution effects
produce a directly measured simulated Tolman slope in the same shallow regime as the JWST result and a volumetric luminosity-density exponent that moves the constant-$\rhoL$ DDR estimator as far towards the non-expanding prediction as the radio data prefer
(see Sec.~\ref{ssec:tng_ladder} and Table~\ref{tab:gamma_aperture} for the quantitative values).
The most significant systematic is the aperture convention with which to analyse the TNG data; we quantify this precisely, finding consistency with the data within $\sim$2$\sigma$.
We also deliver the important clarification that in fact neither the Tolman nor the DDR data lie (unambiguously) on the non-expanding prediction, but rather in between that and the no-evolution FRW prediction. This eliminates the fine-tuning problem that it has been suggested these data pose for FRW cosmology.

The upshot is that the Tolman and angular-size duality results are not two anomalies in need of a coordinated cosmological explanation. They are two measurements exposed to source-population evolution through comparable power-law terms, both of which are naturally of the correct magnitude in the TNG $\Lambda$CDM simulation.
While TNG cannot directly resolve the luminosity evolution of the compact radio sources used in the DDR test, $\alphaDDR$ can be used to infer their volumetric evolution relative to that of the bright galaxies modelled explicitly: we find a relative factor of $\sim1.3$ by $z=2$, consistent with identical evolution within $1\sigma$.

Our work reframes the present Tolman and distance-duality measurements as controlled probes of source evolution and of the systematics that must be calibrated before a clean cosmological inference can be made.
This provides a novel route to inferring the luminosity-density evolution of various kinds of cosmological source, offering new constraints on galaxy formation and evolution at high redshift.

\section*{Acknowledgements}
We thank Amel Durakovic for useful discussions, and the  \texttt{IllustrisTNG} and ASTRODEEP teams for making their products publicly available.
HD is supported by a Royal Society University Research Fellowship (grant no. 211046).
TY acknowledges support from UK Research and Innovation (UKRI) Frontiers Research
Grant [EP/X026639/1], which was selected by the European Research Council (ERC).
RS acknowledges financial support from Science and Technology Facilities Council (STFC) Grant No. ST/X508664/1, the Snell Exhibition of Balliol College, Oxford, and Hintze Fellowship at the Oxford Centre for Astrophysical Surveys, funded through generous support from the Hintze Family Charitable Foundation.
SvH is supported by a Leverhulme Trust International Professorship Grant to S. Sondhi (No. LIP-2020-014).

\section*{Data Availability}
The ASTRODEEP-JWST catalogue is available from the CDS/VizieR archive under identifier \texttt{J/A+A/691/A240}. The \texttt{IllustrisTNG-100} mock images and catalogue products are available from the MAST Illustris HLSP archive at \url{https://archive.stsci.edu/hlsp/illustris}. Code and intermediate products specific to the present analysis are publicly available at \url{https://github.com/harrydesmond/Tolman\_TNG}.

\bibliographystyle{mnras}
\bibliography{refs}

\begin{thebibliography}{}
\makeatletter
\relax
\def\mn@urlcharsother{\let\do\@makeother \do\$\do\&\do\#\do\^\do\_\do\%\do\~}
\def\mn@doi{\begingroup\mn@urlcharsother \@ifnextchar [ {\mn@doi@}
  {\mn@doi@[]}}
\def\mn@doi@[#1]#2{\def\@tempa{#1}\ifx\@tempa\@empty \href
  {http://dx.doi.org/#2} {doi:#2}\else \href {http://dx.doi.org/#2} {#1}\fi
  \endgroup}
\def\mn@eprint#1#2{\mn@eprint@#1:#2::\@nil}
\def\mn@eprint@arXiv#1{\href {http://arxiv.org/abs/#1} {{\tt arXiv:#1}}}
\def\mn@eprint@dblp#1{\href {http://dblp.uni-trier.de/rec/bibtex/#1.xml}
  {dblp:#1}}
\def\mn@eprint@#1:#2:#3:#4\@nil{\def\@tempa {#1}\def\@tempb {#2}\def\@tempc
  {#3}\ifx \@tempc \@empty \let \@tempc \@tempb \let \@tempb \@tempa \fi \ifx
  \@tempb \@empty \def\@tempb {arXiv}\fi \@ifundefined
  {mn@eprint@\@tempb}{\@tempb:\@tempc}{\expandafter \expandafter \csname
  mn@eprint@\@tempb\endcsname \expandafter{\@tempc}}}

\bibitem[\protect\citeauthoryear{{Bassett} \& {Kunz}}{{Bassett} \&
  {Kunz}}{2004}]{BassettKunz2004}
{Bassett} B.~A.,  {Kunz} M.,  2004, \mn@doi [\prd]
  {10.1103/PhysRevD.69.101305}, \href
  {https://ui.adsabs.harvard.edu/abs/2004PhRvD..69j1305B} {69, 101305}

\bibitem[\protect\citeauthoryear{{Bertin} \& {Arnouts}}{{Bertin} \&
  {Arnouts}}{1996}]{SExtractor}
{Bertin} E.,  {Arnouts} S.,  1996, \mn@doi [\aaps] {10.1051/aas:1996164}, \href
  {https://ui.adsabs.harvard.edu/abs/1996A&AS..117..393B} {117, 393}

\bibitem[\protect\citeauthoryear{{Cao}, {Biesiada}, {Jackson}, {Zheng}, {Zhao}
  \& {Zhu}}{{Cao} et~al.}{2017}]{CaoBiesiada2017}
{Cao} S.,  {Biesiada} M.,  {Jackson} J.,  {Zheng} X.,  {Zhao} Y.,   {Zhu}
  Z.-H.,  2017, \mn@doi [\jcap] {10.1088/1475-7516/2017/02/012}, \href
  {https://ui.adsabs.harvard.edu/abs/2017JCAP...02..012C} {2017, 012}

\bibitem[\protect\citeauthoryear{{Conselice}, {Copeland}  \& {Sevillano
  Mu{\~n}oz}}{{Conselice} et~al.}{2026}]{Conselice2026}
{Conselice} C.~J.,  {Copeland} E.~J.,   {Sevillano Mu{\~n}oz} S.,  2026,
  \mn@doi [arXiv e-prints] {10.48550/arXiv.2603.17842}, \href
  {https://ui.adsabs.harvard.edu/abs/2026arXiv260317842C} {p. arXiv:2603.17842}

\bibitem[\protect\citeauthoryear{{Donnari} et~al.,}{{Donnari}
  et~al.}{2019}]{Donnari2019}
{Donnari} M.,  et~al., 2019, \mn@doi [\mnras] {10.1093/mnras/stz712}, \href
  {https://ui.adsabs.harvard.edu/abs/2019MNRAS.485.4817D} {485, 4817}

\bibitem[\protect\citeauthoryear{{Ellis}}{{Ellis}}{2007}]{Ellis2007}
{Ellis} G. F.~R.,  2007, \mn@doi [General Relativity and Gravitation]
  {10.1007/s10714-006-0355-5}, \href
  {https://ui.adsabs.harvard.edu/abs/2007GReGr..39.1047E} {39, 1047}

\bibitem[\protect\citeauthoryear{{Etherington}}{{Etherington}}{1933}]{Etherington1933}
{Etherington} I.~M.~H.,  1933, Philosophical Magazine, \href
  {https://ui.adsabs.harvard.edu/abs/1933PMag...15..761E} {15, 761}

\bibitem[\protect\citeauthoryear{{Fagioli} et~al.,}{{Fagioli}
  et~al.}{2018}]{Fagioli2018}
{Fagioli} M.,  et~al., 2018, \mn@doi [\jcap] {10.1088/1475-7516/2018/11/015},
  \href {https://ui.adsabs.harvard.edu/abs/2018JCAP...11..015F} {2018, 015}

\bibitem[\protect\citeauthoryear{{Fagioli}, {Tortorelli}, {Herbel},
  {Z{\"u}rcher}, {Refregier}  \& {Amara}}{{Fagioli} et~al.}{2020}]{Fagioli2020}
{Fagioli} M.,  {Tortorelli} L.,  {Herbel} J.,  {Z{\"u}rcher} D.,  {Refregier}
  A.,   {Amara} A.,  2020, \mn@doi [\jcap] {10.1088/1475-7516/2020/06/050},
  \href {https://ui.adsabs.harvard.edu/abs/2020JCAP...06..050F} {2020, 050}

\bibitem[\protect\citeauthoryear{{Genel} et~al.,}{{Genel}
  et~al.}{2018}]{Genel2018}
{Genel} S.,  et~al., 2018, \mn@doi [\mnras] {10.1093/mnras/stx3078}, \href
  {https://ui.adsabs.harvard.edu/abs/2018MNRAS.474.3976G} {474, 3976}

\bibitem[\protect\citeauthoryear{{Gurvits}}{{Gurvits}}{1994}]{Gurvits1994}
{Gurvits} L.~I.,  1994, \mn@doi [\apj] {10.1086/173999}, \href
  {https://ui.adsabs.harvard.edu/abs/1994ApJ...425..442G} {425, 442}

\bibitem[\protect\citeauthoryear{{Gurvits}, {Kellermann}  \& {Frey}}{{Gurvits}
  et~al.}{1999}]{GurvitsKellermannFrey1999}
{Gurvits} L.~I.,  {Kellermann} K.~I.,   {Frey} S.,  1999, \mn@doi [\aap]
  {10.48550/arXiv.astro-ph/9812018}, \href
  {https://ui.adsabs.harvard.edu/abs/1999A&A...342..378G} {342, 378}

\bibitem[\protect\citeauthoryear{{Holanda}, {Lima}  \& {Ribeiro}}{{Holanda}
  et~al.}{2010}]{Holanda2010}
{Holanda} R.~F.~L.,  {Lima} J.~A.~S.,   {Ribeiro} M.~B.,  2010, \mn@doi [\apjl]
  {10.1088/2041-8205/722/2/L233}, \href
  {https://ui.adsabs.harvard.edu/abs/2010ApJ...722L.233H} {722, L233}

\bibitem[\protect\citeauthoryear{{Jackson} \& {Jannetta}}{{Jackson} \&
  {Jannetta}}{2006}]{JacksonJannetta2006}
{Jackson} J.~C.,  {Jannetta} A.~L.,  2006, \mn@doi [\jcap]
  {10.1088/1475-7516/2006/11/002}, \href
  {https://ui.adsabs.harvard.edu/abs/2006JCAP...11..002J} {2006, 002}

\bibitem[\protect\citeauthoryear{{Kapahi}}{{Kapahi}}{1987}]{Kapahi1987}
{Kapahi} V.~K.,  1987, in {Hewitt} A.,  {Burbidge} G.,   {Fang} L.~Z.,  eds,
  IAU Symposium Vol. 124, Observational Cosmology. D. Reidel, Dordrecht, pp
  251--265

\bibitem[\protect\citeauthoryear{{Khedekar} \& {Chakraborti}}{{Khedekar} \&
  {Chakraborti}}{2011}]{KhedekarChakraborti2011}
{Khedekar} S.,  {Chakraborti} S.,  2011, \mn@doi [\prl]
  {10.1103/PhysRevLett.106.221301}, \href
  {https://ui.adsabs.harvard.edu/abs/2011PhRvL.106v1301K} {106, 221301}

\bibitem[\protect\citeauthoryear{{Lerner}}{{Lerner}}{2018}]{Lerner2018}
{Lerner} E.~J.,  2018, \mn@doi [\mnras] {10.1093/mnras/sty728}, \href
  {https://ui.adsabs.harvard.edu/abs/2018MNRAS.477.3185L} {477, 3185}

\bibitem[\protect\citeauthoryear{{Li}}{{Li}}{2023}]{Li2023}
{Li} P.,  2023, \mn@doi [\apjl] {10.3847/2041-8213/acdb49}, \href
  {https://ui.adsabs.harvard.edu/abs/2023ApJ...950L..14L} {950, L14}

\bibitem[\protect\citeauthoryear{{Liao}, {Li}, {Cao}, {Biesiada}, {Zheng}  \&
  {Zhu}}{{Liao} et~al.}{2016}]{Liao2016}
{Liao} K.,  {Li} Z.,  {Cao} S.,  {Biesiada} M.,  {Zheng} X.,   {Zhu} Z.-H.,
  2016, \mn@doi [\apj] {10.3847/0004-637X/822/2/74}, \href
  {https://ui.adsabs.harvard.edu/abs/2016ApJ...822...74L} {822, 74}

\bibitem[\protect\citeauthoryear{{Lopez-Corredoira}}{{Lopez-Corredoira}}{2014}]{LopezCorredoira2015}
{Lopez-Corredoira} M.,  2014, in Frontiers of Fundamental Physics 14 (FFP14).
  Proceedings of Science, SISSA, Trieste, p.~85 (\mn@eprint {arXiv}
  {1501.01487}), \mn@doi{10.22323/1.224.0085}

\bibitem[\protect\citeauthoryear{{Lovell}, {Harrison}, {Harikane}, {Tacchella}
  \& {Wilkins}}{{Lovell} et~al.}{2023}]{Lovell2023}
{Lovell} C.~C.,  {Harrison} I.,  {Harikane} Y.,  {Tacchella} S.,   {Wilkins}
  S.~M.,  2023, \mn@doi [\mnras] {10.1093/mnras/stac3224}, \href
  {https://ui.adsabs.harvard.edu/abs/2023MNRAS.518.2511L} {518, 2511}

\bibitem[\protect\citeauthoryear{{Lu}, {Frenk}, {Bose}, {Lacey}, {Cole},
  {Baugh}  \& {Helly}}{{Lu} et~al.}{2025}]{Lu2025}
{Lu} S.,  {Frenk} C.~S.,  {Bose} S.,  {Lacey} C.~G.,  {Cole} S.,  {Baugh}
  C.~M.,   {Helly} J.~C.,  2025, \mn@doi [\mnras] {10.1093/mnras/stae2646},
  \href {https://ui.adsabs.harvard.edu/abs/2025MNRAS.536.1018L} {536, 1018}

\bibitem[\protect\citeauthoryear{{Lubin} \& {Sandage}}{{Lubin} \&
  {Sandage}}{2001a}]{LubinSandage2001b}
{Lubin} L.~M.,  {Sandage} A.,  2001a, \mn@doi [\aj] {10.1086/320401}, \href
  {https://ui.adsabs.harvard.edu/abs/2001AJ....121.2289L} {121, 2289}

\bibitem[\protect\citeauthoryear{{Lubin} \& {Sandage}}{{Lubin} \&
  {Sandage}}{2001b}]{LubinSandage2001c}
{Lubin} L.~M.,  {Sandage} A.,  2001b, \mn@doi [\aj] {10.1086/322133}, \href
  {https://ui.adsabs.harvard.edu/abs/2001AJ....122.1071L} {122, 1071}

\bibitem[\protect\citeauthoryear{{Lubin} \& {Sandage}}{{Lubin} \&
  {Sandage}}{2001c}]{LubinSandage2001d}
{Lubin} L.~M.,  {Sandage} A.,  2001c, \mn@doi [\aj] {10.1086/322134}, \href
  {https://ui.adsabs.harvard.edu/abs/2001AJ....122.1084L} {122, 1084}

\bibitem[\protect\citeauthoryear{{Marinacci} et~al.,}{{Marinacci}
  et~al.}{2018}]{Marinacci2018}
{Marinacci} F.,  et~al., 2018, \mn@doi [\mnras] {10.1093/mnras/sty2206}, \href
  {https://ui.adsabs.harvard.edu/abs/2018MNRAS.480.5113M} {480, 5113}

\bibitem[\protect\citeauthoryear{{Marshall} et~al.,}{{Marshall}
  et~al.}{2022}]{Marshall2022}
{Marshall} M.~A.,  et~al., 2022, \mn@doi [\mnras] {10.1093/mnras/stac2111},
  \href {https://ui.adsabs.harvard.edu/abs/2022MNRAS.516.1047M} {516, 1047}

\bibitem[\protect\citeauthoryear{{Merlin} et~al.,}{{Merlin}
  et~al.}{2024}]{Merlin2024}
{Merlin} E.,  et~al., 2024, \mn@doi [\aap] {10.1051/0004-6361/202451409}, \href
  {https://ui.adsabs.harvard.edu/abs/2024A&A...691A.240M} {691, A240}

\bibitem[\protect\citeauthoryear{{Naiman} et~al.,}{{Naiman}
  et~al.}{2018}]{Naiman2018}
{Naiman} J.~P.,  et~al., 2018, \mn@doi [\mnras] {10.1093/mnras/sty618}, \href
  {https://ui.adsabs.harvard.edu/abs/2018MNRAS.477.1206N} {477, 1206}

\bibitem[\protect\citeauthoryear{{Nelson} et~al.,}{{Nelson}
  et~al.}{2018}]{Nelson2018}
{Nelson} D.,  et~al., 2018, \mn@doi [\mnras] {10.1093/mnras/stx3040}, \href
  {https://ui.adsabs.harvard.edu/abs/2018MNRAS.475..624N} {475, 624}

\bibitem[\protect\citeauthoryear{{Nelson} et~al.,}{{Nelson}
  et~al.}{2019}]{NelsonHLSP}
{Nelson} D.,  et~al., 2019, \mn@doi [Computational Astrophysics and Cosmology]
  {10.1186/s40668-019-0028-x}, \href
  {https://ui.adsabs.harvard.edu/abs/2019ComAC...6....2N} {6, 2}

\bibitem[\protect\citeauthoryear{{Pahre}, {Djorgovski}  \& {de
  Carvalho}}{{Pahre} et~al.}{1996}]{Pahre1996}
{Pahre} M.~A.,  {Djorgovski} S.~G.,   {de Carvalho} R.~R.,  1996, \mn@doi
  [\apjl] {10.1086/309872}, \href
  {https://ui.adsabs.harvard.edu/abs/1996ApJ...456L..79P} {456, L79}

\bibitem[\protect\citeauthoryear{{Pedregosa} et~al.,}{{Pedregosa}
  et~al.}{2011}]{Pedregosa2011}
{Pedregosa} F.,  et~al., 2011, \mn@doi [Journal of Machine Learning Research]
  {10.48550/arXiv.1201.0490}, \href
  {https://ui.adsabs.harvard.edu/abs/2011JMLR...12.2825P} {12, 2825}

\bibitem[\protect\citeauthoryear{{Pillepich} et~al.,}{{Pillepich}
  et~al.}{2018a}]{Pillepich2018Model}
{Pillepich} A.,  et~al., 2018a, \mn@doi [\mnras] {10.1093/mnras/stx2656}, \href
  {https://ui.adsabs.harvard.edu/abs/2018MNRAS.473.4077P} {473, 4077}

\bibitem[\protect\citeauthoryear{{Pillepich} et~al.,}{{Pillepich}
  et~al.}{2018b}]{Pillepich2018}
{Pillepich} A.,  et~al., 2018b, \mn@doi [\mnras] {10.1093/mnras/stx3112}, \href
  {https://ui.adsabs.harvard.edu/abs/2018MNRAS.475..648P} {475, 648}

\bibitem[\protect\citeauthoryear{{Pillepich} et~al.,}{{Pillepich}
  et~al.}{2019}]{Pillepich2019}
{Pillepich} A.,  et~al., 2019, \mn@doi [\mnras] {10.1093/mnras/stz2338}, \href
  {https://ui.adsabs.harvard.edu/abs/2019MNRAS.490.3196P} {490, 3196}

\bibitem[\protect\citeauthoryear{{Planck Collaboration} et~al.,}{{Planck
  Collaboration} et~al.}{2016}]{Planck2016}
{Planck Collaboration} et~al., 2016, \mn@doi [\aap]
  {10.1051/0004-6361/201525830}, \href
  {https://ui.adsabs.harvard.edu/abs/2016A&A...594A..13P} {594, A13}

\bibitem[\protect\citeauthoryear{{Sandage} \& {Lubin}}{{Sandage} \&
  {Lubin}}{2001}]{LubinSandage2001a}
{Sandage} A.,  {Lubin} L.~M.,  2001, \mn@doi [\aj] {10.1086/320394}, \href
  {https://ui.adsabs.harvard.edu/abs/2001AJ....121.2271S} {121, 2271}

\bibitem[\protect\citeauthoryear{{Snyder}, {Pe{\~n}a}, {Yung}, {Rose},
  {Kartaltepe}  \& {Ferguson}}{{Snyder} et~al.}{2023}]{Snyder2022}
{Snyder} G.~F.,  {Pe{\~n}a} T.,  {Yung} L.~Y.~A.,  {Rose} C.,  {Kartaltepe} J.,
    {Ferguson} H.,  2023, \mn@doi [\mnras] {10.1093/mnras/stac3397}, \href
  {https://ui.adsabs.harvard.edu/abs/2023MNRAS.518.6318S} {518, 6318}

\bibitem[\protect\citeauthoryear{{Springel}}{{Springel}}{2010}]{Springel2010}
{Springel} V.,  2010, \mn@doi [\mnras] {10.1111/j.1365-2966.2009.15715.x},
  \href {https://ui.adsabs.harvard.edu/abs/2010MNRAS.401..791S} {401, 791}

\bibitem[\protect\citeauthoryear{{Springel} et~al.,}{{Springel}
  et~al.}{2018}]{Springel2018}
{Springel} V.,  et~al., 2018, \mn@doi [\mnras] {10.1093/mnras/stx3304}, \href
  {https://ui.adsabs.harvard.edu/abs/2018MNRAS.475..676S} {475, 676}

\bibitem[\protect\citeauthoryear{{Tolman}}{{Tolman}}{1930}]{Tolman1930}
{Tolman} R.~C.,  1930, \mn@doi [Proceedings of the National Academy of Science]
  {10.1073/pnas.16.7.511}, \href
  {https://ui.adsabs.harvard.edu/abs/1930PNAS...16..511T} {16, 511}

\bibitem[\protect\citeauthoryear{{Tolman}}{{Tolman}}{1934}]{Tolman1934}
{Tolman} R.~C.,  1934, {Relativity, Thermodynamics, and Cosmology}.
Clarendon Press, Oxford

\bibitem[\protect\citeauthoryear{{Tsymbal}, {Raikov}  \& {Lovyagin}}{{Tsymbal}
  et~al.}{2026}]{TolmanJWST}
{Tsymbal} V.~V.,  {Raikov} A.~A.,   {Lovyagin} N.~Y.,  2026, \mn@doi [arXiv
  e-prints] {10.48550/arXiv.2604.27867}, \href
  {https://ui.adsabs.harvard.edu/abs/2026arXiv260427867T} {p. arXiv:2604.27867}

\bibitem[\protect\citeauthoryear{{Uzan}, {Aghanim}  \& {Mellier}}{{Uzan}
  et~al.}{2004}]{Uzan2004}
{Uzan} J.-P.,  {Aghanim} N.,   {Mellier} Y.,  2004, \mn@doi [\prd]
  {10.1103/PhysRevD.70.083533}, \href
  {https://ui.adsabs.harvard.edu/abs/2004PhRvD..70h3533U} {70, 083533}

\bibitem[\protect\citeauthoryear{{Vogelsberger} et~al.,}{{Vogelsberger}
  et~al.}{2020}]{Vogelsberger2020}
{Vogelsberger} M.,  et~al., 2020, \mn@doi [\mnras] {10.1093/mnras/staa137},
  \href {https://ui.adsabs.harvard.edu/abs/2020MNRAS.492.5167V} {492, 5167}

\bibitem[\protect\citeauthoryear{{Weinberger} et~al.,}{{Weinberger}
  et~al.}{2017}]{Weinberger2017}
{Weinberger} R.,  et~al., 2017, \mn@doi [\mnras] {10.1093/mnras/stw2944}, \href
  {https://ui.adsabs.harvard.edu/abs/2017MNRAS.465.3291W} {465, 3291}

\bibitem[\protect\citeauthoryear{{Whitney}, {Conselice}, {Bhatawdekar}  \&
  {Duncan}}{{Whitney} et~al.}{2019}]{Whitney2020}
{Whitney} A.,  {Conselice} C.~J.,  {Bhatawdekar} R.,   {Duncan} K.,  2019,
  \mn@doi [\apj] {10.3847/1538-4357/ab53d4}, \href
  {https://ui.adsabs.harvard.edu/abs/2019ApJ...887..113W} {887, 113}

\bibitem[\protect\citeauthoryear{{Wilkins} et~al.,}{{Wilkins}
  et~al.}{2023}]{Wilkins2023}
{Wilkins} S.~M.,  et~al., 2023, \mn@doi [\mnras] {10.1093/mnras/stac3280},
  \href {https://ui.adsabs.harvard.edu/abs/2023MNRAS.519.3118W} {519, 3118}

\bibitem[\protect\citeauthoryear{{Yung} et~al.,}{{Yung}
  et~al.}{2022}]{Yung2022}
{Yung} L.~Y.~A.,  et~al., 2022, \mn@doi [\mnras] {10.1093/mnras/stac2139},
  \href {https://ui.adsabs.harvard.edu/abs/2022MNRAS.515.5416Y} {515, 5416}

\bibitem[\protect\citeauthoryear{{Yung}, {Somerville}, {Finkelstein}, {Wilkins}
   \& {Gardner}}{{Yung} et~al.}{2024}]{Yung2024}
{Yung} L.~Y.~A.,  {Somerville} R.~S.,  {Finkelstein} S.~L.,  {Wilkins} S.~M.,
  {Gardner} J.~P.,  2024, \mn@doi [\mnras] {10.1093/mnras/stad3484}, \href
  {https://ui.adsabs.harvard.edu/abs/2024MNRAS.527.5929Y} {527, 5929}

\makeatother
\end{thebibliography}

\bsp
\label{lastpage}
\end{document}